\documentstyle[12pt]{article}

\catcode`\@=11

\newcount\hour
\newcount\minute
\newtoks\amorpm \hour=\time\divide\hour by 60\minute
=\time{\multiply\hour by 60 \global\advance\minute by-\hour}
\edef\standardtime{{\ifnum\hour<12 \global\amorpm={am}%
        \else\global\amorpm={pm}\advance\hour by-12 \fi
        \ifnum\hour=0 \hour=12 \fi
        \number\hour:\ifnum\minute<10
        0\fi\number\minute\the\amorpm}}
\edef\militarytime{\number\hour:\ifnum\minute<10
0\fi\number\minute}

\def\draftlabel#1{{\@bsphack\if@filesw {\let\thepage\relax
   \xdef\@gtempa{\write\@auxout{\string
      \newlabel{#1}{{\@currentlabel}{\thepage}}}}}\@gtempa
   \if@nobreak \ifvmode\nobreak\fi\fi\fi\@esphack}
        \gdef\@eqnlabel{#1}}
\def\@eqnlabel{}
\def\@vacuum{}
\def\marginnote#1{}
\def\draftmarginnote#1{\marginpar{\raggedright\scriptsize\tt#1}}
\def\draft{
        \pagestyle{plain}
        \overfullrule=2pt
        \oddsidemargin -.5truein
        \def\@oddhead{\sl \phantom{\today\quad\militarytime} \hfil
        \smash{\Large\sl DRAFT} \hfil \today\quad\militarytime}
        \let\@evenhead\@oddhead
        \let\label=\draftlabel
        \let\marginnote=\draftmarginnote
        \def\ps@empty{\let\@mkboth\@gobbletwo
        \def\@oddfoot{\hfil \smash{\Large\sl DRAFT} \hfil}
        \let\@evenfoot\@oddhead}
        \def\@eqnnum{(\theequation)\rlap{\kern\marginparsep\tt\@eqnlabel}%
        \global\let\@eqnlabel\@vacuum}  }

\newcommand{\rf}[1]{(\ref{#1})}
\renewcommand{\theequation}{\thesection.\arabic{equation}}
\renewcommand{\thefootnote}{\fnsymbol{footnote}}
\newcommand{\newsection}{    % Numeration of eqs. is automatic
\setcounter{equation}{0}\section}
\def\appendix#1{\addtocounter{section}{1}\setcounter{equation}{0}
\renewcommand{\thesection}{\Alph{section}}
\section*{Appendix \thesection\protect\indent \parbox[t]{11.15cm}{#1}}
\addcontentsline{toc}{section}{Appendix \thesection\ \ \ #1}}

\def\be{\begin{equation}}
\def\ee{\end{equation}}

\def\PP{{\cal P}}

\def\x'{\mathaccent 19 x}
\def\th'{\mathaccent 19 \theta}
\def\barth'{\mathaccent 19 {\bar{\theta}}}
\def\phx{\mathaccent 19 {\phantom{x}} }

\def\ipr{{i^\prime}}
\def\jpr{{j^\prime}}
\def\kpr{{k^\prime}}
\def\lpr{{l^\prime}}

\begin{document}

%\draft

\begin{flushright}
hep-th/0112044
\\
FIAN/TD/01-17
\end{flushright}

\vspace{1cm}

\begin{center}

{\Large\bf  Type IIB Green-Schwarz superstring in \\

\medskip
plane wave Ramond-Ramond background}

\vspace{2.5cm} R.R. Metsaev

\vspace{1cm} {\it Department of Theoretical Physics, P.N. Lebedev
Physical Institute,

\medskip
Leninsky prospect 53, 119991, Moscow, Russia}

\vspace{3.5cm} {\bf Abstract}

\end{center}

We construct the covariant $\kappa$-symmetric superstring action
for type $IIB$ superstring on plane wave space supported by
Ramond-Ramond background. The action is defined as a $2d$
sigma-model on the coset superspace. We fix the fermionic and
bosonic light-cone gauges in the covariant Green-Schwarz
superstring action and find the light-cone string Lagrangian and
the Hamiltonian. The resulting light-cone gauge action is
quadratic in both the bosonic and fermionic superstring $2d$
fields, and therefore, this model can be explicitly quantized. We
also obtain a realization of the generators of the basic
superalgebra in terms of the superstring $2d$ fields in the
light-cone gauge.

\newpage

\renewcommand{\thefootnote}{\arabic{footnote}}
\setcounter{footnote}{0}

\newsection{Introduction}

Until recently precisely two maximally supersymmetric solutions
of the $IIB$ supergravity were known. The first solution is the
flat ten dimensional Minkowski space (and its toroidal
compactification) and the second solution is the famous
$AdS_5\times S^5$ space supported by Ramond-Ramond  charges
\cite{schwarz}-\cite{sw}, which plays a distinguished role in the
dual description of gauge theories. Surprisingly, very recently a
new maximally supersymmetric solution of $IIB$ supergravity was
found \cite{blau} that  turns out to be a ten dimensional plane
wave space supported by the RR 5-form flux. This new solution is a
ten dimensional counterpart of the solution found for eleven
dimensional supergravity in \cite{kg1} (see also \cite{kg2}).

Applying arguments similar to those in \cite{kraj}, one can expect
that as in the case of the $AdS_5\times S^5$ space, the plane wave
RR background is an exact solution not only of the equations of
motion of $IIB$ supergravity but also of the equations of motion
for massless modes of the type $IIB$ superstring. The action of
the $AdS_5\times S^5$ superstring constructed in \cite{mt1} (see
also \cite{krr}-\cite{mtt}) turns out to be very complicated for
explicit quantization. In this paper, motivated by desire to find
explicitly quantizable superstring model in curved target space
time supported by the Ramond-Ramond flux, we investigate classical
mechanics of the superstring propagating in the plane wave
Ramond-Ramond background. We find that the light-cone action of
this superstring model is quadratic in bosonic as well as
fermionic superstring $2d$ fields and can therefore be quantized
in a rather straightforward way. The interest in plane wave RR
background also comes from the fact that this relatively simple
model may serve as a training ground for the study of a more
interesting case of superstring in the $AdS_5\times
S^5$/RR-charge background.

One possible approach to the construction of the action is to
start with the general type $IIB$ superstring action in \cite{ghm}
and plug in the type $IIB$ superfields \cite{hw} representing the
plane wave RR background. This approach, however, is indirect and
somewhat complicated because it does not explicitly use the basic
symmetries of the problem. Our strategy is instead to use the
basic superalgebra underlying the symmetry of the plane wave RR
background. As in the previous construction of the type IIB
superstring action in $AdS_5 \times S^5$ \cite{mt1}, we obtain the
space-time supersymmetric and $\kappa$-symmetric action in terms
of the invariant Cartan one-forms defined on the appropriate
coset superspace. The supercoset construction turned out to be
very effective and fruitful and was used to construct $AdS(2)$ and
$AdS(3)$ superstring actions \cite{zho}-\cite{parrey}, brane
actions in super $AdS/RR$ backgrounds \cite{mt2},\cite{daw}, and
various $AdS$ supermembrane actions \cite{wpps}-\cite{dik}.

This paper is organized as follows.

In section \ref{algebra}, we describe the structure of the basic
symmetry superalgebra of the plane wave RR background and the
invariant Cartan 1-forms on the coset superspace $(x,\theta)$.

In section \ref{action}, we present the covariant superstring
action on the plane wave RR background in the coordinate free
form, i.e. in terms of the Cartan 1-forms. As in the cases of
flat and $AdS_5\times S^5$ spaces, it is given by the sum of the
`kinetic' or `Nambu' term ($2d$ integral of the quadratic term in
Cartan 10-beins) and Wess-Zumino term ($3d$ integral of a closed
3-form ${\cal H}$ on the superspace) with the coefficient of the
WZ term fixed by the requirement of $\kappa$-symmetry (Siegel
symmetry).

In section \ref{2dform}, we find the explicit $2d$ form of the
action by choosing a specific WZ parametrization of the coset
superspace. As in the $AdS_5\times S^5$ space, the resulting
action is given by the covariantization of the flat-space GS
action plus terms containing higher powers of the fermionic
coordinates $\theta$.

In section \ref{Lcaction}, we evaluate Cartan 1-forms in
fermionic light-cone gauge, which we use to derive the superstring
action in the fermionic $\kappa$-symmetry light-cone gauge. After
this, we fix the $2d$ diffeomorphism symmetry by choosing bosonic
light-cone gauge. The resulting light-cone action turns out to be
quadratic in bosonic and fermionic $2d$ string fields. This
allows us to solve the equations of motion explicitly. We discuss
the superstring action in various parametrization of the plane
wave background and in the WZ and the Killing parametrizations of
superspace.

Section \ref{Phaction} is devoted to the  light-cone phase space
approach to superstring theory. We  fix the analog of the GGRT
\cite{ggrt} bosonic light-cone gauge  and derive  the phase space
analog of the  superstring Lagrangian of  Section 5 and the
corresponding light-cone gauge  Hamiltonian.

In Section \ref{Ncharges}, we obtain a realization of the
generators of the basic symmetry superalgebra as Noether charges
expressed in terms of  the $2d$ fields that are coordinates of
the plane wave superstring in the light-cone gauge.

Section \ref{concl} summarizes our conclusions. Some  technical
details are collected in two Appendices. In Appendix A, we
summarize our  notation  and definitions. In Appendix B, we
explain the construction of dynamical supercharges realized in
terms of $2d$ superstring fields.

\newsection{Symmetry Superalgebra of plane wave
RR \\ background }\label{algebra}

We start with reviewing the plane wave RR background of the type
$IIB$ supergravity and its symmetry superalgebra \cite{blau}. The
bosonic sector of $IIB$ supergravity includes the graviton, a
complex scalar, a complex two-index antisymmetric tensor  and a
real four-index antisymmetric tensor whose five-index field
strength is (anit)self-dual. The line element of the plane wave
background given by

\be\label{lin1}
 ds^2=2dx^+dx^- - m^2 x_I^2 dx^{+2} + dx_I^2\,, \ee
$I=1,\ldots 8$, describes ten dimensional lorentzian symmetric
space. Here $m$ is a dimensionful parameter and $x^+$ is taken to
be light-cone evolution parameter. This metric is supported by
non-vanishing four-index antisymmetric RR field whose five-index
field strength takes  the values

\be\label{fiestr1} F^{-i_1\ldots i_4} = 2m \epsilon^{i_1\ldots
i_4}\,,\qquad \qquad F^{-\ipr_1\ldots \ipr_4} = 2m
\epsilon^{\ipr_1\ldots \ipr_4}\,,\ee $i,j=1,\ldots 4$,
$\ipr,\jpr=5,\ldots 8$. All the remaining bosonic fields (as well
as fermionic fields) are set to be equal zero. Surprisingly, this
background has 32 Killing spinors and therefore preserves  32
supersymmetries. This can easily be checked by considering the
equations for Killing spinors ${\cal D}\epsilon =0$ with the
covariant differential \footnote{ We adopt the normalization in
which the Einstein equations take the form \\ $R^{\mu\nu} = (1/24)
F^{\mu \rho_1\ldots \rho_4} F^{\nu\rho_1\ldots \rho_4}$ and take
the five-index field strength to be anti-self dual:
$F^{\mu_1\ldots \mu_5}= - (1/120)\epsilon^{\mu_1\ldots
\mu_5\rho_1\ldots \rho_5} F^{\rho_1\ldots \rho_5}$,
$\epsilon^{01\ldots 9} = 1$\,. The expression for ${\cal D}$ in
\rf{covder1} is restricted to 16-component $\epsilon$ that is a
half of a 32-component positive chirality spinor. The indices
$\mu,\nu,\rho=0,1,\ldots 9$ are $so(9,1)$ vector indices, i.e.
tangent space indices. In light cone frame these indices take
values $+,-,I$ where $I=i,\ipr$.}

\be\label{covder1} {\cal D}  = d +\frac{1}{4}\omega^{\mu\nu}
\gamma^{\mu\nu} - \frac{\rm i}{960} \gamma^{\mu_1\ldots \mu_5}
\bar{\gamma}^\mu e^\mu F^{\mu_1\ldots \mu_5}\,, \ee where $e^\mu$
are 10-beins of plane wave space \rf{lin1} and $\omega^{\mu\nu}$
is the corresponding Lorentz connection. Inserting the above
expressions for five-index field strength \rf{fiestr1} in
\rf{covder1}, one verifies that the integrability condition of the
Killing spinors equations, i.e. ${\cal D}^2=0$, is satisfied
automatically. Because all the other conditions are also satisfied
automatically, there are 32 Killing spinors. In what follows,
unless otherwise specified, we set

\be m=1\,\ee to simplify our expressions.

 We now describe the
basic symmetry superalgebra of the plane wave RR background
\rf{lin1},\rf{fiestr1}. The even (bosonic) part of the
superalgebra includes ten translation generators $P^\mu$, the
$SO(4)$ rotation generators $J^{ij}$, $i,j=1,\ldots, 4$, the
$SO^\prime(4)$ rotation generators $J^{\ipr\jpr}$,
$\ipr,\jpr=5,\ldots, 8$ and eight rotation generators in the
$(x^-,x^I)$ plane $J^{+I}$, $I=1,\ldots,8$. The odd (fermionic)
part of the superalgebra consists of the complex 16-component
spinor $Q_\alpha$, $\alpha=1,\ldots 16$, which is half of
32-component  negative chirality spinor

Commutation relations between the even generators are given by

\be [P^-,P^I] = -J^{+I}\,,\ee

\be [P^I,J^{+J}] = -\delta^{IJ}P^+\,, \qquad
[P^-,J^{+I}]=P^I\,,\ee

\be [P^i,J^{jk}] = \delta^{ij}P^k -\delta^{ik}P^j\,, \qquad
[P^\ipr,J^{\jpr\kpr}] = \delta^{\ipr\jpr}P^\kpr
-\delta^{\ipr\kpr}P^\jpr\,,\ee

\be [J^{+i},J^{jk}] = \delta^{ij}J^{+k} -\delta^{ik}J^{+j}\,,
\qquad [J^{+\ipr},J^{\jpr\kpr}] = \delta^{\ipr\jpr}J^{+\kpr}
-\delta^{\ipr\kpr}J^{+\jpr}\,,\ee

\be [J^{ij},J^{kl}] = \delta^{jk}J^{il} + 3\hbox{ terms}\,,
\qquad [J^{\ipr\jpr},J^{\kpr\lpr}] = \delta^{\jpr\kpr}J^{\ipr\lpr}
+ 3\hbox{ terms}\,. \ee

Commutation relations  between the even and odd parts are

 \be\label{jq1} [J^{ij},Q_\alpha] =
 \frac{1}{2}Q_\beta(\gamma^{ij})^\beta{}_\alpha
\,, \qquad [J^{\ipr\jpr},Q_\alpha] =
\frac{1}{2}Q_\beta(\gamma^{\ipr\jpr})^\beta{}_\alpha \,, \ee

 \be [J^{+I},Q_\alpha] =
 \frac{1}{2}Q_\beta(\gamma^{+I})^\beta{}_\alpha\,,
\ee

\be [P^\mu,Q_\alpha] =  \frac{\rm i}{2} Q_\beta( \Pi
\gamma^+\bar{\gamma}^\mu)^\beta{}_\alpha\,,
 \ee together with the commutators that follow from these by complex
 conjugation.  Here, $\gamma^\mu$ are $16\times 16$ gamma matrices,
$\gamma^{+I}=
 \gamma^{[+}\bar{\gamma}^{I]}$, and
$\Pi$ is the product of four gamma matrices (see Appendix A for
details).

The anticommutator takes the form

\begin{eqnarray} \{Q_\alpha, \bar{Q}_\beta \}
&=& -2{\rm i}\gamma^\mu_{\alpha\beta} P^\mu - 2
(\bar{\gamma}^i\Pi)_{\alpha\beta}J^{+i} - 2 (\bar{\gamma}^\ipr
\Pi^\prime)_{\alpha\beta}J^{+\ipr}
\nonumber\\[7pt]
\label{qq} &+&(\bar{\gamma}^+\gamma^{ij}\Pi)_{\alpha\beta}J^{ij}
+(\bar{\gamma}^+\gamma^{\ipr\jpr}\Pi^\prime)_{\alpha\beta}J^{\ipr\jpr}\,.
\end{eqnarray}
All the other commutators and anticommutators vanish. The bosonic
generators are assumed to be antihermitean while the fermionic
generators are conjugated to each other, $\bar{Q}_\alpha =
(Q_\alpha)^\dagger$. The generators of the full transformation
group $G$ of the plane wave RR superspace  are $P^\mu$,
$Q_\alpha$, $\bar{Q}_\alpha$, $J^{ij}$, $J^{\ipr\jpr}$, $J^{+I}$.
The generators of the stability subgroup $H$ are $J^{ij}$,
$J^{\ipr\jpr}$, $J^{+I}$. The plane wave RR superspace is defined
then as coset superspace $G/H$.

A few remarks are in order.

(i) The (anti)commutation relations of the superalgebra are
invariant under $U(1)$ transformation of supercharges:
$Q\rightarrow e^{{\rm i}\phi}Q$, $\bar{Q}\rightarrow e^{-{\rm
i}\phi}\bar{Q}$. This $U(1)$ symmetry reflects the fact that the
plane wave RR background respects the original $U(1)$ symmetry of
$IIB$ supergravity.

(ii) In contrast to flat and $AdS_5\times S^5$ cases, the above
superalgebra does not include rotation generator in the
$(x^+,x^-)$ plane, usually denoted by $J^{+-}$, and the rotation
generators in the $(x^+,x^I)$ planes which are $J^{-I}$. We note
that for the bosonic string in a plane wave NS-NS background, it
is the absence of $J^{-I}$ generators that explains why the
critical dimension cannot be obtained from the usual operator
formalism argument \cite{forhor}. The interesting fact
\cite{blau} is that the dimension of the bosonic subalgebra of the
superalgebra under consideration, which is equal to 30, coincides
with the dimension of the isometry algebra of the $AdS_5 \times
S^5$ space.

(iii)
 The dimensionful parameter of the plane wave geometry can be
introduced by rescaling the generators as $ P^\mu \rightarrow
P^\mu/m^2$, $Q_\alpha  \rightarrow Q_\alpha/m$. The limit as $m
\rightarrow 0$ then gives the subalgebra of $d=10$, $IIB$
Poincar\'e superalgebra.

\subsection{Cartan  1-forms   }
%%%%%%%%%%%%%%%%%%%%%%%%%%%%%%%%%%%%%%%%%%%

To find the super-invariant and
 $\kappa$-invariant  string action we use  the formalism of
Cartan forms defined on the coset superspace.\footnote{For some
applications of the formalism of  the Cartan forms  on coset
superspaces see \cite{vann}.} The left-invariant Cartan 1-forms

\be L^A = dX^M L^A_{M}\ ,  \qquad \ \ \ X^M=(x^\mu,
\theta^\alpha,\bar{\theta}^\alpha)\ee are given by

\be\label{expan1} G^{-1}dG = L^\mu P^\mu + L^\alpha\bar{Q}_\alpha
+ \bar{L}^\alpha Q_\alpha + \frac{1}{2}L^{\mu\nu}J^{\mu\nu}\,,\ee
where we impose the conditions

\be\label{lpc} L^{+\mu}=0\,, \qquad L^{i\jpr} = 0\,,\ee which
simply reflect the fact that the generators $J^{-\mu}$ and
$J^{i\jpr}$ are not included in superalgebra. $L^\mu$ are the
10-beins, $L^\alpha$, $\bar{L}^\alpha$, ($\bar{L}^\alpha =
(L^\alpha)^\dagger$) are the two spinor 16-beins, and
$L^{\mu\nu}$ are the Cartan $H$ connections.\footnote{ $L^\alpha$
is a half of 32-component  positive chirality spinor.} They
satisfy the Maurer-Cartan equations implied by the structure of
the superalgebra,

\begin{eqnarray}
\label{mc1} && dL^\mu = -L^{\mu\nu}L^\nu -2{\rm i}\bar{L}
\bar{\gamma}^\mu L\,,
\\
\label{mc2} && dL^\alpha =
-\frac{1}{4}L^{\mu\nu}(\gamma^{\mu\nu})^\alpha{}_\beta L^\beta
+\frac{\rm i}{2}L^\mu(\Pi\gamma^+\bar{\gamma}^\mu)^\alpha{}_\beta
L^\beta\,,
\\
 \label{mc3} &&
 d\bar{L}^\alpha =
-\frac{1}{4}L^{\mu\nu}(\gamma^{\mu\nu})^\alpha{}_\beta
\bar{L}^\beta -\frac{\rm
i}{2}L^\mu(\Pi\gamma^+\bar{\gamma}^\mu)^\alpha{}_\beta
\bar{L}^\beta\,. \end{eqnarray}  We note that we suppress  the
exterior products symbols for 1-forms and use the following sign
conventions under permutations of Cartan 1-forms:

\be \label{sigc} L^\mu L^\nu = - L^\nu L^\mu\,, \qquad L^\mu
L^\alpha = - L^\alpha L^\mu\,,\qquad L^\alpha L^\beta=L^\beta
L^\alpha\,.\ee

The dependence on the dimensionful parameter $m$ can be restored
by simultaneously  rescaling the Cartan 1-forms and coordinates as
$L^\mu \rightarrow m^2 L^\mu$, $L^{\mu\nu} \rightarrow
L^{\mu\nu}$, $L^\alpha \rightarrow mL^\alpha$, $x^\mu\rightarrow
m^2 x^\mu$, $\theta \rightarrow m\theta$.

For comparison, we note that  in  the flat superspace case,

\be \label{supf} G(x,\theta) = \exp( x^\mu P^\mu  + L^\alpha
\bar{Q}_\alpha +\bar{L}^\alpha Q_\alpha)\,,\ee

\be [P^\mu, P^\nu]=0\,, \ \ \ \{ Q_\alpha, \bar{Q}_\beta\} =
-2{\rm i} \gamma_{\alpha\beta}^\mu P^\mu \ee and  thus  the coset
space (super)vielbeins  in $G^{-1} d G = L^A T_A$ are therefore
given by

\be \label{flat} L_0^\mu = d x^\mu  - {\rm i} \bar \theta
\bar{\gamma}^\mu d\theta - {\rm i}\theta \bar{\gamma}^\mu
d\bar{\theta}\,, \qquad L_0 = d \theta \ .
 \ee

\newsection{Superstring action as sigma model on $G/H$ coset
superspace}\label{action}

In this section, closely following the general method suggested in
\cite{mt1}, we construct the superstring action that satisfies the
following conditions (some of which are not completely
independent):

a) its bosonic part is the standard $\sigma$-model with the
plane wave geometry as a  target

\ \ \ space;

b) it  has global super-invariance with respect to supersymmetry
algebra above

\ \ \ described;

c)  it is invariant under  the local $\kappa$-symmetry;

d)  it reduces to the  standard   Green-Schwarz type IIB
superstring action in the

\ \ \   flat-space ($m \rightarrow 0$) limit.

\noindent As in \cite{mt1} we find   that  such an action exists
and is {\it unique}.   Its leading $\theta^2$ fermionic term
contains the  required  coupling to the RR 5-form field
background.

It is useful to recall that the flat-space GS superstring
Lagrangian \cite{gs} can be written in the  manifestly
supersymmetric form in terms of (super)vielbeins  (\ref{flat}) as
a sum of the `kinetic' term and the WZ term \cite{hm},

\be \label{lag0} {\cal L}_0 = {\cal L}_{0\, kin} +{\cal
L}_{0\,WZ}\,,\ee where the kinetic and WZ terms are given
by\footnote{We  use Minkowski signature $2d$ world-sheet metric
$g_{ab}$ with $g\equiv - \det g_{ab}$.}

\be\label{gs1} {\cal L}_{0\,kin} = -\frac{1}{2}\sqrt{g}g^{ab}
L_{0b}^\mu L_{0a}^\mu\,, \qquad {\cal L}_{0\,WZ} = d^{-1}{\cal
H}_0\,,\ee and $L_{0a}^A =\partial_a X^M L_{0 M}^A$. The 3-form
${\cal H}_0$ being closed and exact allows manifestly
supersymmetric representation\footnote{ For fermionic coordinates
we assume the convention $(\theta_1 \theta_2)^\dagger =
\bar{\theta}_2 \bar{\theta}_1$, $\theta_1\theta_2
=-\theta_2\theta_1$, while for fermionic Cartan 1-forms we adopt
$(L_1 \wedge L_2)^\dagger = -\bar{L}_2 \wedge \bar{L}_1$,
$L_1\wedge L_2  = L_2\wedge L_1$.}

\be\label{flath} {\cal H}_0 ={\rm i} L_0^\mu L_0\bar{\gamma}^\mu
L_0 +h.c.\ee The coefficient of the WZ term is fixed by the
condition of the local $\kappa$-invariance \cite{gs}. Using the
explicit representation for (super)vielbeins (\ref{flat}) one
observes that the  3-form in the WZ term is indeed exact and thus
finds the explicit $2d$ form of the WZ part of the superstring
Lagrangian \cite{gs}

\be \label{gsexp}{\cal L}_{WZ} = - {\rm i}\epsilon^{ab}(\partial_a
x^\mu -\frac{\rm i}{2}\bar{\theta}\bar{\gamma}^\mu \partial_a
\theta -\frac{\rm i}{2}\theta\bar{\gamma}^\mu\partial_a
\bar{\theta})\theta \bar{\gamma}^\mu\partial_b \theta + h.c.\,,
\ee which is invariant under global supersymmetry only up to a
total derivative. The action that we find below is the
generalization of \rf{lag0}  to the case  where the  free bosonic
term is replaced by the sigma model on the plane wave RR
background.

We now turn to superstring in the plane wave RR background. As in
the flat space, the Lagrangian is given by the sum of
$\sigma$-model term ${\cal L}_{kin}$ and the WZ term ${\cal
L}_{WZ}=d^{-1}{\cal H}$. To satisfy invariance with respect to
the symmetry superalgebra, both ${\cal L}_{kin}$ and ${\cal H}$
should be constructed in terms of the Cartan 1-forms $L^\mu$ and
$L^\alpha$. The basic observation is that under the action of an
arbitrary element of the group $G$ these forms transform as
tangent vectors (spinors) of the stability group $H$. Therefore
any invariant of stability group constructed in terms of $L^\mu$
and $L^\alpha$ is automatically invariant under full
transformations of $G$.

The structure of ${\cal L}_{kin}$ is fixed by conditions (a) and
(b) and can be obtained from ${\cal L}_{0kin}$ in \rf{gs1} by
replacing $L_0^\mu$ with the Cartan 1-forms of the plane wave RR
background $L^\mu$. As to the WZ part, it turns out that the only
{\it relevant} {\it closed} 3-form built out of $L^\mu$,
$L^\alpha$ that is invariant  under transformations of
$SO(4)\otimes SO^\prime(4)$ and the those generated by the
light-cone boost generators $J^{+I}$ is given by

\be\label{hcom}
 {\cal H} = {\cal H}^q + {\cal H}^{\bar{q}},\qquad
 {\cal H}^q =  ({\cal H}^{\bar{q}})^\dagger\,,
 \ee
where

 \be\label{hq}
 {\cal H}^q  = {\rm i} L^\mu L \bar{\gamma}^\mu L\,.
 \ee
The fact that this form is indeed closed can be demonstrated as
follows. Using  Maurer-Cartan equations (\ref{mc1})-(\ref{mc3}),
we find terms in $d{\cal H}^q$ that are proportional to the
(super)vielbeins $L^\mu$, $L^\alpha$, and the Cartan $H$
connections $L^{\mu\nu}$. Taking the relation

\be \bar{\gamma}^\mu \gamma^{\nu\rho} =
\bar{\gamma}^{\mu\nu\rho}+\eta^{\mu\nu}\bar{\gamma}^\rho -
\eta^{\mu\rho}\bar{\gamma}^\nu \ee into account and using that
$(\bar{\gamma} ^{\mu\nu\rho})_{\alpha\beta}$ are antisymmetric in
$\alpha,\beta$ we find that the terms proportional to
$L^{\mu\nu}$ cancel out. The remaining part of $d{\cal H}^q$ is
then given by

\be d{\cal H}^q = L^\mu L^\nu L^\alpha (\bar{\gamma}^\mu \Pi
\gamma^+ \bar{\gamma}^\nu)_{\alpha\beta}L^\beta\,. \ee Using that

\be (\bar{\gamma}^i\Pi)_{\alpha\beta}, \quad
(\Pi\gamma^+\bar{\gamma}^{ij})_{\alpha\beta}, \quad
(\Pi\gamma^+\bar{\gamma}^{\ipr\jpr})_{\alpha\beta} \hbox{ are
antisymmetric in } (\alpha,\beta) \ee together with the symmetry
properties of the products of Cartan 1-forms given in
(\ref{sigc}), we find that $d{\cal H}^q =0$. Because all Cartan
1-forms of plane wave RR background reduce to flat Cartan 1-forms
in the flat-space limit the 3-form ${\cal H}$ \rf{hcom},\rf{hq}
also reduces to the 3-form in the GS action (\ref{flath}). As in
flat space, the value of the
 overall coefficient in front ${\cal H}$  is fixed to be 1
by the requirement of $\kappa$-symmetry of the whole action
 (which is proved below).
The final expression for  the Lagrangian written
 in the manifestly invariant
form  in terms of the (super)vielbeins $L^\mu$ and $L^\alpha$
 thus has the same structure as  the GS Lagrangian (\ref{gs1}),

\be\label{mylag} {\cal L} = {\cal L}_{kin} + {\cal L}_{WZ}\,,\ee

\be\label{my} {\cal L}_{kin} = -\frac{1}{2}\sqrt{g}g^{ab}
L_{b}^\mu L_{a}^\mu\,, \qquad {\cal L}_{WZ} = d^{-1}{\cal H}\ee
and, indeed, reduces to    (\ref{gs1}) in the flat-space limit.
Because the 3-form  ${\cal H}$  is closed it can be  represented
as ${\cal H}= d {\cal B}$ in a local coordinate system; the
action then takes the  usual $2d$ sigma-model form, which is
considered in what follows.

\subsection{The $\kappa$-symmetry invariance}

The action (\ref{mylag}) is invariant with respect to the  local
$\kappa$-transformations \cite{gs},\cite{ws}. These can be
conveniently written in terms of the Cartan 1-form and the
variations defined by

\be \widehat{\delta x}{}^\mu \equiv  {\delta X}^M L_M^\mu\,,
\qquad \widehat{\delta x}{}^{\mu\nu} \equiv  {\delta X}^M
L_M^{\mu\nu}\,, \qquad \widehat{\delta \theta}{}^\alpha \equiv
{\delta X}^M L_M^\alpha\,. \ee To formulate the
$\kappa$-transformations we introduce a complex 16-component
spinor $(\kappa^a)_\alpha$ (the corresponding 32-component
spinor  has negative chirality) that is a $2d$ vector on the
world-sheet. The $\kappa$-transformation then takes the form

\be \widehat {\delta x}{}^\mu =0\,,\qquad\widehat {\delta \theta}
= 2L_a^\mu \gamma^\mu \kappa^a\,,\ee

\be \delta (\sqrt{g} g^{ab}) = - 8{\rm i}\sqrt{g} (L^a
\bar{\kappa}^b + L^b\bar{\kappa}^a -\frac{1}{2}g^{ab} L_c
\bar{\kappa}^c) +h.c.\,, \ee where $\bar{\kappa} =
\kappa^\dagger$. The $\kappa$-transformation parameter satisfies
the  (anti) self duality constraints which in complex notation
that we use take the form\footnote{The selfduality constraints
(\ref{sdcon}) are counterpart of standard constraints formulated
in terms of two Majorana Weyl spinors $\kappa^1$, $\kappa^2$:
$\epsilon^{ab}\kappa_b^1 = - \sqrt{g}\kappa^{a 1}$,
$\epsilon^{ab}\kappa_b^2 = \sqrt{g}\kappa^{a 2}$.}

\be\label{sdcon} \frac{\epsilon^{ab}}{\sqrt{g}}\ \kappa_b =
-\bar{\kappa}^a\,, \qquad \frac{\epsilon^{ab}}{\sqrt{g}}\
\bar{\kappa}_b = -\kappa^a \,.\ee

To demonstrate the $\kappa$-invariance we use the following
expressions for the variations of the Cartan 1-forms:

\begin{eqnarray}
\delta L^\mu  &=& d\widehat{\delta x}{}^\mu + L^\nu
\widehat{\delta x}{}^{\nu\mu} + L^{\mu\nu} \widehat{\delta
x}{}^\nu
 +  2{\rm i} \bar{L}\bar{\gamma}^\mu \widehat{\delta \theta}
- 2{\rm i} \widehat{\delta\bar{\theta}}\bar{\gamma}^\mu L\,,
\\
\delta L & = & d \widehat{\delta \theta} - \frac{\rm i}{2}L^\mu
\Pi \gamma^+\bar{\gamma}^\mu \widehat{\delta \theta}
+\frac{1}{4}L^{\mu\nu} \gamma^{\mu\nu} \widehat{\delta \theta}
\nonumber\\
&+&\frac{\rm i}{2} \widehat{\delta x}{}^\mu \Pi \gamma^+
\bar{\gamma}^\mu L -\frac{1}{4}\widehat{\delta x}{}^{\mu\nu}
\gamma^{\mu\nu} L\,.
\end{eqnarray}
The crucial  relation that allows us to check the
$\kappa$-invariance of the superstring action directly in terms of
the Cartan 1-forms is

\be \delta {\cal H}^q =d\Lambda^q\,,\qquad \Lambda^q =
  {\rm i}\widehat{\delta x}{}^\mu
L \bar{\gamma}^\mu L + 2{\rm i} L^\mu L \bar{\gamma}^\mu
\widehat{\delta \theta}\,. \ee

%%%%%%%%%%%%%%%%%%%%%%%%%%%%%%%%%%%%%%%%%%%%%%%%%%%%%
\newsection{ Explicit 2-dimensional  form of the
action}\label{2dform}
%%%%%%%%%%%%%%%%%%%%%%%%%%%%%%%%%%%%%%%%%%%%%%%
To find the  explicit form of the WZ part of the superstring
Lagrangian in terms of the coordinate $2d$ field $\theta$ which
generalizes (\ref{gsexp}) we choose a particular parametrization
of the coset representative $G$ in (\ref{expan1}):

\be\label{gxt} G(x,\theta) = g(x) {\rm g} (\theta)\ , \ \ \ \ \ \
\ \ \ {\rm g} (\theta) =\exp(\theta^\alpha \bar{Q}_\alpha
+\bar{\theta}^\alpha Q_\alpha)\,. \ee Here, $g(x)$ is a coset
representative of $G_{bos}/H$, i.e., $x=x^\mu$ provides a certain
parametrization of the plane wave geometry (which we do not need
to  specify  in this section). We note that choosing the coset
representative in form \rf{gxt} corresponds to the Wess-Zumino
type gauge in the plane wave/RR superspace, while another, $
G(x,\theta) = {\rm g} (\theta) g(x)$, corresponds to the Killing
gauge. These ``gauges"    (better to be called
``parametrizations") do not reduce the number of the fermionic
degrees of freedom but only specialize a  choice  of fermionic
coordinates. The covariant action given in this Section
corresponds to the  WZ parametrization.

To represent the WZ term in (\ref{my}) as an density over the
2-dimensional space we  use the standard  trick of rescaling
$\theta\rightarrow \theta_t\equiv t\theta$,

\be {\cal L}_{\rm WZ} =  {\cal L}_{\rm WZ}(t=1) \,, \qquad \ \
{\cal L}_{WZ}(t) = d^{-1}({\cal H}^q_t +h.c.)\,, \qquad {\cal
H}_t^q = {\cal H}^q (\theta_t) \ . \ee We then have the obvious
relation

\be
\label{hdec9} {\cal H}_{t=1}^q = {\cal H}_{t=0}^q + \int_0^1
dt \,
\partial_t {\cal H}_t^q\,. \ee
Inserting $G_t \equiv G(x,t\theta)$ given by (\ref{gxt}) in
definition of Cartan 1-forms and setting $L_t^A = L^A(x,t\theta)$
we obtain the equations for the `shifted' Cartan one-forms $L_t^A$

\begin{eqnarray}
\label{t1} &&
\partial_t L_t =  d\theta +
\frac{1}{4}L_t^{\mu\nu}\gamma^{\mu\nu} \theta -\frac{\rm
i}{2}L_t^\mu \Pi\gamma^+\bar{\gamma}^\mu \theta\,,
\\[3pt]
\label{t2} &&
\partial_t L_t^\mu
=   - 2{\rm i}\theta  \bar{\gamma}^\mu \bar{L}_t -  2{\rm
i}\bar{\theta} \bar{\gamma}^\mu L_t\,,
\\[3pt]
\label{t3} &&
\partial_t L_t^{-i} =  2\theta \bar{\gamma}^i\Pi
\bar{L}_t -  2\bar{\theta} \bar{\gamma}^i\Pi L_t\,,
\\[3pt]
&&
\partial_t L_t^{-\ipr} =  2\theta \bar{\gamma}^\ipr\Pi^\prime
\bar{L}_t - 2\bar{\theta} \bar{\gamma}^\ipr\Pi^\prime L_t \,,
\\[3pt]
\label{t5} &&
\partial_t L_t^{ij} =  - 2\theta\bar{\gamma}^+
\gamma^{ij}\Pi \bar{L}_t
 + 2\bar{\theta} \bar{\gamma}^+ \gamma^{ij}\Pi L_t\,,
\\
\label{t6} && \partial_t L_t^{\ipr\jpr} =  - 2\theta
\bar{\gamma}^+ \gamma^{\ipr\jpr}\Pi^\prime \bar{L}_t
 + 2\bar{\theta} \bar{\gamma}^+ \gamma^{\ipr\jpr}\Pi^\prime L_t\,.
 \end{eqnarray}
These equations should be supplemented by the initial conditions

\be\label{inicon} L_{t=0}^\mu = e^\mu\,,\qquad L_{t=0}^{\mu\nu}=
\omega^{\mu\nu}\,,\qquad L_{t=0}=0\,,\ee $\omega^{+\mu}=0$,
$\omega^{i\jpr}$=0. Here $e^\mu$ are the 10-beins of the plane
wave geometry and $\omega^{\mu\nu}$ is the corresponding Lorentz
connection. Using these differential equations we prove that

\be
\partial_t {\cal H}_t^q = -2{\rm i}d(L_t^\mu \theta \bar{\gamma}^\mu
L_t)\,. \ee Taking into account that ${\cal H}_{t=0}^q=0$, we
obtain the desired representation for ${\cal L}_{WZ}$,

 \be \label{wz1} {\cal L}_{WZ} = -2{\rm i}
\int_0^1 dt\, L_t^\mu\, \theta \bar{\gamma}^\mu L_t +h.c. \ee To
summarize the complete superstring Lagrangian is given by the sum
of the kinetic part \rf{my} and the $2d$ form of the WZ part
\rf{wz1},

\be\label{comlag} {\cal L} = -\frac{1}{2}\sqrt{g}g^{ab} L_{b}^\mu
L_{a}^\mu -2{\rm i} \int_0^1 dt\, \epsilon^{ab} L_{at}^\mu\,
(\theta \bar{\gamma}^\mu L_{b t} + \bar{\theta}\bar{\gamma}^\mu
\bar{L}_{b t})\,. \ee

In the next section we use this representation to derive the
light-cone gauge superstring action.

As a side remark we note that the equations for the Cartan 1-forms
can be solved in a rather straightforward way. For this, we
collect the fermionic Cartan 1-forms and $\theta'$s in two
vectors\footnote{These ${\bf L}$ and $\Theta$ should not be
confused with 32-component spinors.}

\be {\bf L} =\left(\begin{array}{c} L \\[7pt]
\bar{L}\end{array}\right)\,,\qquad
\Theta =\left(\begin{array}{c} \theta \\[7pt]
\bar{\theta}\end{array}\right)\,.\ee The solution to equations
\rf{t1}-\rf{t6} and initial conditions \rf{inicon} then leads to
the Cartan 1-forms $L= L_{t=1}$ given by

 \be\label{exprep1} {\bf L} = \frac{\hbox{sihn} {\cal
M}}{\cal M}{\cal D}\Theta\,,\ee

\be \label{exprep2}  L^\mu = e^\mu -2{\rm
i}\bar{\Theta}\,\bar{\gamma}^\mu \frac{\hbox{cosh}{\cal
M}-1}{{\cal M}^2}{\cal D}\Theta\,,\ee where the covariant
derivative ${\cal D}$ is\footnote{This is essentially the same
derivative which appeared in the Killing spinor equations of
$IIB$ supergravity\rf{covder1}.}

 \be {\cal D}\Theta =
(d +\frac{1}{4} \omega^{\mu\nu} \gamma^{\mu\nu} -\frac{\rm i}{2}
e^\mu \Pi \gamma^+ \bar{\gamma}^\mu \sigma_3) \Theta\,. \ee The
square of the covariant differential ${\cal D}$ is equal to zero,
${\cal D}^2=0$, and the Killing spinor equation ${\cal D}
\epsilon =0$ is therefore integrable. The block $32\times 32$
matrix ${\cal M}^2$ is given by\footnote{Action of the asterisk
on the product of fermions in \rf{mmat} is defined to be
$(F_1F_2)^* = \bar{F}_1\bar{F}_2$.}

\be\label{mmat} ({\cal M}^2)^\alpha{}_\beta
=\left(\begin{array}{cc}
 A^\alpha{}_\beta   &  B^\alpha{}_\beta
 \\[7pt]
 -(B^\alpha{}_\beta)^* & -(A^\alpha{}_\beta)^*
 \end{array}\right)\,,
 \ee
where the $16\times 16$ matrices $A$ and $B$ are given by

\begin{eqnarray}
 A^\alpha{}_\beta & =& -(\Pi\gamma^+\bar{\gamma}^\mu\theta)^\alpha
(\bar{\theta}\bar{\gamma}^\mu)_\beta -(\gamma^{+i}\theta)^\alpha
(\bar{\theta}\bar{\gamma}^i\Pi)_\beta -
(\gamma^{+\ipr}\theta)^\alpha
(\bar{\theta}\bar{\gamma}^\ipr\Pi^\prime)_\beta
\nonumber\\
&+& \frac{1}{2}(\gamma^{ij}\theta)^\alpha
(\bar{\theta}\bar{\gamma}^+\gamma^{ij}\Pi)_\beta +
\frac{1}{2}(\gamma^{\ipr\jpr}\theta)^\alpha
(\bar{\theta}\bar{\gamma}^+\gamma^{\ipr\jpr}\Pi^\prime)_\beta\,,
 \\
B^\alpha{}_\beta & =& -(\Pi\gamma^+\bar{\gamma}^\mu\theta)^\alpha
(\theta\bar{\gamma}^\mu)_\beta +(\gamma^{+i}\theta)^\alpha
(\theta\bar{\gamma}^i\Pi)_\beta + (\gamma^{+\ipr}\theta)^\alpha
(\bar{\theta}\bar{\gamma}^\ipr\Pi^\prime)_\beta
\nonumber\\
&-& \frac{1}{2}(\gamma^{ij}\theta)^\alpha
(\theta\bar{\gamma}^+\gamma^{ij}\Pi)_\beta -
\frac{1}{2}(\gamma^{\ipr\jpr}\theta)^\alpha
(\theta\bar{\gamma}^+\gamma^{\ipr\jpr}\Pi^\prime)_\beta\,.
\end{eqnarray}

%%%%%%%%%%%%%%%%%%%%%%%%%%%%%%%%%%%%%%%%%%%%%%%%%%%
\newsection{Light-cone superstring action
in plane wave  R-R background }\label{Lcaction}
%%%%%%%%%%%%%%%%%%%%%%%%%%%%%%%%%%%%%%%%%%%

In this Section we find the form of the type $IIB$ superstring
action in the plane wave background with R-R 5-form flux in the
light-cone gauge. Our discussion  of the light-cone gauge fixing
closely repeats the same steps as in Refs.\cite{mt3,mtt}, where
the $AdS_5 \times S^5$ case was treated.

In flat space  the superstring light-cone gauge  fixing procedure
consists of the two  stages:

(I) fermionic light-cone gauge choice, i.e., fixing the
$\kappa$-symmetry by  $\bar{\gamma}^+ \theta=0$

(II) bosonic light-cone gauge choice, i.e., using   the conformal
gauge $\sqrt {g} g^{ab} =\eta^{ab}$ and fixing  the  residual
conformal diffeomorphism symmetry by $x^+(\tau,\sigma)  =  p^+
\tau$.

Our fermionic  $\kappa$-symmetry light-cone gauge is the same as
in flat superstring  $\bar{\gamma}^+ \theta =0$. One usually
imposes the $\kappa$-symmetry light-cone gauge by starting with
the explicit representation for the superstring Lagrangian
\rf{comlag} in terms of $\theta'$s. However it is convenient to
first impose the light-cone gauge at the level of the Cartan forms
$L^\mu$, $L^\alpha$ and then to use them in \rf{comlag}. In what
follows we adopt this strategy.\footnote{This strategy was used in
\cite{pes},\cite{kr} while deriving $\kappa$-symmetry fixed
action for long superstring in $AdS_5\times S^5$ and for
$\kappa$-symmetry fixed light-cone $AdS$ superstring in
\cite{mt3}.}

%%%%%%%%%%%%%%%%%%%%%%%%%%%%%%%%%%%%%%%%%%
\subsection{Cartan 1-forms in Wess-Zumino parametrization in
\\
fermionic light-cone gauge}
%%%%%%%%%%%%%%%%%%%%%%%%%%%%%%%%%%%%%%%%%%%%%%%%%%%%%%%%%%%%%%%%%

We first consider fixing the fermionic $\kappa$-symmetry in the
Cartan 1-forms written  in the  WZ parametrization. The fermionic
$\kappa$-symmetry light-cone gauge is defined as

\be\label{ferg} \bar{\gamma}^+ \theta = \bar{\gamma}^+
\bar{\theta} = 0\,. \ee To simplify our expressions, we choose
the parametrization of the plane wave space such that the
 bosonic bodies of the Cartan $H$ connections
$L^{ij}$, $L^{\ipr\jpr}$ (defined in \rf{inicon}) are equal to
zero,

\be\label{spepar1} \omega^{ij} =0\,,\qquad
\omega^{\ipr\jpr}=0\,.\ee The $\kappa$-symmetry gauge fixed
bosonic Cartan 1-forms $L^\mu$ are then found to be

\be \label{lpi} L^+ = e^+\,, \qquad L^I = e^I\,, \ee

\be \label{lmin} L^- = e^-  - {\rm
i}(\bar{\theta}\bar{\gamma}^-d\theta+ \theta\bar{\gamma}^-
d\bar{\theta}) -2e^+ \bar{\theta}\bar{\gamma}^- \Pi \theta \,,\ee
while the fermionic 16-beins $L^\alpha$ take the form

\be \label{lfer} L = d\theta - {\rm i}e^+ \Pi \theta\,. \ee

These expressions for the Cartan 1-forms are valid for an
arbitrary parametrization of the bosonic 10-beins $e^\mu$
satisfying  relations \rf{spepar1}. In what follows, we use the
10-beins\footnote{Note that such 10-beins can be obtained by
using the following coset representative of bosonic body in
\rf{gxt}: $g(x) = \exp(x^+ P^-) \exp( x^- P^+ + x^I P^I)$.}

\be\label{boscar} e^+ = dx^+ \,,\qquad e^- = dx^-
-\frac{1}{2}x_I^2 dx^+\,,\qquad e^I = dx^I\,.\ee

Expressions \rf{lpi}-\rf{lfer} can obviously be obtained from the
general solution for the Cartan 1-forms given in
\rf{exprep1},\rf{exprep2}. However the most convenient procedure
is to derive them from differential equations
(\ref{t1}),(\ref{t6}) using the fermionic $\kappa$-symmetry
light-cone gauge (\ref{ferg}). We now outline this procedure.

First of all, using \rf{ferg} in \rf{t2} for $\mu=+$, we find the
equation $\partial_t L_t^+=0$, which together with \rf{inicon}
gives

\be L_t^+ = e^+ \,.\ee Second, multiplying equations \rf{t1} by
$\bar{\gamma}^+$ and taking \rf{ferg} and \rf{lpc} into account,
we find the equation $\partial_t (\bar{\gamma}^+ L_t )=0$, which
together with \rf{inicon} gives

\be\label{ltc} \bar{\gamma}^+ L_t =0\,. \ee Now inserting the
decomposition of unity $1=(\gamma^+\bar{\gamma}^- +
\gamma^-\bar{\gamma}^+)/2$  between $L_t$ and $\theta$ in the
r.h.s. of \rf{t2} and taking  \rf{ferg} and \rf{ltc} into account
we obtain the equation $\partial_t L_t^I =0$, which together with
\rf{inicon} leads to

\be L_t^I = e^I\,. \ee Third,  using \rf{ferg} in \rf{t5},\rf{t6}
gives the equations $\partial_t L_t^{ij}=\partial_t
L^{\ipr\jpr}=0$. Because of \rf{spepar1} these equations imply

 \be L_t^{ij} = 0\,,\qquad L_t^{\ipr\jpr} =0\,. \ee
Using of these relations and gauge \rf{ferg} in equations \rf{t1}
gives

\be
\partial_t L_t = d\theta -{\rm i}e^+ \Pi \theta\,.
\ee  We thus find

\be L_t = t(d\theta -{\rm i}e^+ \Pi \theta)\,. \ee Inserting this
$L_t$ in the equation for $L_t^-$, we obtain

\be \partial_t L_t^- = -2{\rm
i}t(\bar{\theta}\bar{\gamma}^-d\theta+ \theta\bar{\gamma}^-
d\bar{\theta} ) -4te^+ \bar{\theta}\bar{\gamma}^- \Pi \theta\,.
\ee

The solution of this equations for $t=1$ gives the expression for
$L^-$ in \rf{lmin}.

%%%%%%%%%%%%%%%%%%%%%%%%%%%%%%%%%%%%%%%%%%%%%%%%%%%%%%%%%%%%%%%%%
\subsection{$\kappa$-symmetry gauge fixed superstring
action \\ in WZ parametrization}
%%%%%%%%%%%%%%%%%%%%%%%%%%%%%%%%%%%%%%%%%%%%%%%%%%%%%%%%%%%%%%%%%%%

Because we use coset parametrization \rf{gxt} the
$\kappa$-symmetry gauge fixed action given in this Section
corresponds to the WZ parametrization of superspace. The
light-cone gauge action in the Killing parametrization is
discussed in Section \ref{kilpar}.

Inserting the above expressions for Cartan 1-forms
\rf{lpi}-\rf{lfer} into Lagrangian \rf{comlag} and using
\rf{boscar} we obtain the $\kappa$-symmetry gauge fixed
superstring Lagrangian  in terms of the light-cone supercoset
coordinates

\be \label{ltot2} {\cal L}  = {\cal L}_{kin} + {\cal L}_{WZ}\,,\ee
where the kinetic and WZ parts are given by

\begin{eqnarray}\label{kin2}
{\cal L}_{kin} & = &-\frac{1}{2}\sqrt{g}g^{ab}\Bigl(2\partial_a
x^+
\partial_b x^- -x_I^2 \partial_a x^+\partial_b x^+ + \partial_a
x^I \partial_b x^I\Bigr)
\nonumber\\[7pt]
& - & {\rm i}\sqrt{g}g^{ab}\partial_b x^+\Bigl(
\bar{\theta}\bar{\gamma}^-\partial_a\theta + \theta\bar{\gamma}^-
\partial_a\bar{\theta} + 2{\rm i}\partial_a x^+
\bar{\theta}\bar{\gamma}^- \Pi \theta\Bigr)\,,
\end{eqnarray}

\be\label{wz2} {\cal L}_{WZ} = {\rm i}\epsilon^{ab}\partial_a x^+
\theta \bar{\gamma}^-\partial_b \theta +h.c. \ee The kinetic
terms in \rf{kin2} can be obtained in a straightforward way. To
find the WZ part we make rescaling $\theta \rightarrow t\theta$
in \rf{lpi}-\rf{lfer} and then integrate over $t$ in \rf{comlag}.
Note that in deriving representations for ${\cal L}_{kin}$,
${\cal L}_{WZ}$ in \rf{kin2},\rf{wz2} we made the redefinition
$x^\mu \rightarrow -x^\mu$. This form of the Lagrangian is most
convenient for deriving the Noether charges (see Section
\ref{Ncharges}).

The superstring Lagrangian given by equations \rf{ltot2}-\rf{wz2}
can obviously be rewritten as

 \be\label{ltot3}
{\cal L}={\cal L}_{B}+ {\cal L}_{F}\,. \ee Here, ${\cal L}_B =
-\frac{1}{2} \sqrt{g} g^{ab}  G_{\mu\nu}(x) \partial_a x^\mu
\partial_b x^\nu$ given in the first line in \rf{kin2} is the
standard bosonic sigma model with the plane wave geometry as the
target space and ${\cal L}_F$ given by the second line in
\rf{kin2} and by \rf{wz2} is the fermionic part of the Lagrangian

\begin{eqnarray}
\label{lagfer} {\cal L}_F &=&
 -  {\rm i}\sqrt{g}g^{ab}\partial_b x^+\Bigl(
\bar{\theta}\bar{\gamma}^-\partial_a\theta + \theta\bar{\gamma}^-
\partial_a\bar{\theta} + 2{\rm i}\partial_a x^+
\bar{\theta}\bar{\gamma}^- \Pi \theta\Bigr) \nonumber\\
&+&{\rm i}\epsilon^{ab}\partial_a x^+ (\theta
\bar{\gamma}^-\partial_b \theta +\bar{\theta}
\bar{\gamma}^-\partial_b \bar{\theta})\,.\end{eqnarray}

%%%%%%%%%%%%%%%%%%%%%%%%%%%%%%%%%%%%%%%%%%%%%%%%%%%%%%%%%%
 \subsection{Bosonic light-cone gauge and ``$2d$  spinor"  form  of the
action  }
%%%%%%%%%%%%%%%%%%%%%%%%%%%%%%%%%%%%%%%%%%

As in  the flat space case \cite{gs}, the resulting action can
then be put into  the ``2-d spinor" form.  Indeed, the 8+8
fermionic degrees of freedom  can be organized into 8 Majorana
$2d$ spinors, defined  in  {\it flat} $2d$ geometry. This form of
the action may be useful in establishing the relation to the NSR
formulation.

As in the flat  case, we should eliminate the $\partial x^+
$-factors from the  kinetic
 terms of fermionic fields \rf{lagfer}.
 In flat space this  was  possible by choosing the  bosonic
 light-cone
 gauge.    In the  BDHP formulation \cite{bdh,poly}
 that we are using, this can be  done by fixing the conformal
 gauge as

   \be\label{conf}
 \sqrt{g} g^{ab}=\eta^{ab}\,,\qquad -\eta^{00} = \eta^{11} =1  \ee
     and then noting
   that because the resulting equation
   $(\partial_0^2-\partial_1^2) x^+=0$ has the general solution
   $x^+ (\tau,\sigma) = f( \tau-\sigma) + h( \tau+ \sigma)$
we can fix the residual conformal diffeomorphism symmetry
   on the plane by choosing

   \be \label{xpsol} x^+(\tau,\sigma)  = \tau\,,\ee
   where we put $p^+=1$. Using \rf{conf} and \rf{xpsol} in ${\cal L}_B$
and ${\cal
   L}_F$ \rf{ltot3},\rf{lagfer} we find the bosonic and fermionic
   Lagrangians

\begin{eqnarray}\label{lagbos2}
&&  {\cal L}_B  = -\frac{1}{2}\partial_a x^I
\partial^a x^I - \frac{1}{2} x_I^2\,,
\\
\label{lagfer2} && {\cal L}_F   =  {\rm i} (
\bar{\theta}\bar{\gamma}^-\partial_0\theta +
\theta\bar{\gamma}^-  \partial_0\bar{\theta}+ \theta
\bar{\gamma}^-\partial_1 \theta + \bar{\theta}
\bar{\gamma}^-\partial_1 \bar{\theta})
 - 2 \bar{\theta}\bar{\gamma}^- \Pi \theta\,.
\end{eqnarray} Similarly to the flat space case the fermionic
Lagrangian ${\cal L}_F$ can be rewritten as

\be{\cal L}_F = - {\rm i} \bar{\psi}\bar{\gamma}^-\varrho^a
\partial_a \psi
 +{\rm i} \bar{\psi} \bar{\gamma}^- \Pi \psi\,.\ee
Here $\varrho^a$ are $2\times 2$ Dirac matrices

\be\varrho^0 \equiv -{\rm i}\sigma_2\,,\qquad \varrho^1\equiv
\sigma_1\,,\ee $\bar{\psi}\equiv \psi^T\varrho^0$, $\psi^T$
denotes transposition of $2d$ spinor and $\psi'$s are related to
the ($2d$ scalar) fermionic fields $\theta'$s by\footnote{In our
notation $\bar{\psi}\varrho^a
\partial_a \psi = -\psi^1(\partial_0 +
\partial_1)\psi^1 - \psi^2(\partial_0 - \partial_1)\psi^2
$.}

\be \psi = \left(\begin{array}{c} \psi^1\\[7pt]
\psi^2\end{array}\right)\,,\qquad \psi^1 \equiv \theta^1\,,\qquad
\psi^2 \equiv\theta^2 \,,\ee where $\theta^1$ and $\theta^2$ are
real spinors related to the original complex spinors $\theta$ and
$\bar{\theta}$ by \rf{comrea}. From this, it is clear that
$\psi'$s are $8$ Majorana $2d$ spinors.

If we restore a dependence on the dimensionful parameter $m$
\rf{lin1} then we obtain

\be\label{lag2dform}
 {\cal L}  = -\frac{1}{2}\partial_a x^I
\partial^a x^I - \frac{m^2}{2} x_I^2
- {\rm i} \bar{\psi}\bar{\gamma}^-\varrho^a \partial_a \psi
 +{\rm i}m \bar{\psi} \bar{\gamma}^- \Pi \psi\,. \ee
Thus the total superstring Lagrangian describes 8 free massive
$2d$ scalars and $8$ free  massive Majorana $2d$ fermionic fields
propagating in a flat two dimensional world-sheet. This
superstring model can obviously be quantized in a rather
straightforward way.

%%%%%%%%%%%%%%%%%%%%%%%%%%%%%%%%%%%%%%%%%%%%%%%%%%%%%%%%%%
 \subsection{Solution to superstring equations of motion }
%%%%%%%%%%%%%%%%%%%%%%%%%%%%%%%%%%%%%%%%%%

In contrast to the $AdS_5\times S^5$ case the superstring in the
plane wave Ramond-Ramond background has free equations of motion
that can be easily solved. To analyze the equations of motion we
use Lagrangian formulated in terms of
$(x^I,\,\theta,\,\bar{\theta})$ (see
\rf{ltot3},\rf{lagbos2}\rf{lagfer2}):

\begin{eqnarray}\label{lag4}
 {\cal L}  & =&  -\frac{1}{2}\partial_a x^I
\partial^a x^I - \frac{1}{2} x_I^2
\nonumber\\
& +&  {\rm i} ( \bar{\theta}\bar{\gamma}^-\partial_0\theta +
\theta\bar{\gamma}^-  \partial_0\bar{\theta}+ \theta
\bar{\gamma}^-\partial_1 \theta + \bar{\theta}
\bar{\gamma}^-\partial_1 \bar{\theta})
 - 2 \bar{\theta}\bar{\gamma}^- \Pi \theta\,.
 \end{eqnarray}
The equations of motion and periodicity conditions for bosonic
fields are given by

\be (-\partial_0^2 +\partial_1^2)x^I -x^I=0\,,\ee

\be x^I(\sigma+1,\tau) = x^I(\sigma,\tau)\,,\qquad 0\leq \sigma
\leq 1\,. \ee The solution to these equations can be written as

\be \label{xIsol} x^I(\sigma,\tau) = \cos \tau\, x_0^I + \sin
\tau\, p_0^I + {\rm i}\sum_{n=\!\!\!\!/\,\, 0
}\frac{1}{\omega_n}\Bigl( \varphi_n^1(\sigma,\tau) \alpha_n^{1I}
+ \varphi_n^2(\sigma,\tau) \alpha_n^{2I}\Bigr)\,, \ee where the
$x_0^I$ and $p_0^I$ are zero modes, while $\alpha_n^{1I}$ and
$\alpha_n^{2I}$ are string oscillators modes. The base functions
$\varphi^{1,2}_n(\sigma,\tau)$ for the right and left movers are
given by

\be\varphi^1_n(\sigma,\tau) = \exp(-{\rm i}(\omega_n \tau
-k_n\sigma))\,,\qquad \varphi^2_n(\sigma,\tau) = \exp(-{\rm
i}(\omega_n \tau + k_n\sigma))\,,\ee where the frequencies
$\omega_n$ are defined by

\be \omega_n =\sqrt{k_n^2 + 1},\quad n>0;\qquad
 \omega_n
=-\sqrt{k_n^2 + 1},\quad  n < 0;\ee

\be k_n = 2\pi n\,,\qquad n =\pm 1,\pm 2 \ldots  \ee

Using the rules implied by standard canonical Hamiltonian approach
(which is systematically developed in Section \ref{secphaact}) we
obtain the Poisson brackets

\be\label{aacom} [p_0^I,x_0^J]=\delta^{IJ}\,,\qquad
[\alpha_m^{1I},\alpha_n^{1J}] =\frac{\rm i}{2}\omega_m
\delta_{m+n,0}\delta^{IJ}\,, \qquad [\alpha_m^{2I},\alpha_n^{2J}]
=\frac{\rm i}{2}\omega_m \delta_{m+n,0}\delta^{IJ}\,.\ee

We now turn to fermionic fields. Lagrangian \rf{lag4} gives the
equations of motion for $\theta'$s

 \be
\partial_0 \theta +\partial_1 \bar{\theta} +{\rm i}\Pi \theta
=0\,, \qquad \partial_0 \bar{\theta} + \partial_1 \theta  -{\rm
i}\Pi \bar{\theta} =0\,,\ee that should be supplemented by the
periodicity conditions
$\theta(\sigma+1,\tau)=\theta(\sigma,\tau)$. We prefer to
re-formulate these equations in terms of two {\it real} fermionic
fields $\theta^1$ and $\theta^2$ (see \rf{comrea})

 \be \label{freeq1}(\partial_0 +\partial_1) \theta^1  -
\Pi \theta^2 =0\,,\ee

\be \label{freeq2}(\partial_0 - \partial_1) \theta^2  + \Pi
\theta^1 =0\,.\ee Solutions to these equations are found to be

\be \label{th1sol}\theta^1(\sigma,\tau) =\cos\tau\, \theta_0^1
+\sin\tau \Pi\theta_0^2+ \sum_{n=\!\!\!\!/\,\, 0}c_n\Bigl(
\varphi_n^1(\sigma,\tau) \theta^1_n + {\rm
i}(\omega_n-k_n)\varphi_n^2(\sigma,\tau)\Pi \theta_n^2\Bigr)\,,\ee

\be \label{th2sol} \theta^2(\sigma,\tau) =\cos\tau\, \theta_0^2 -
\sin\tau \Pi\theta_0^1+ \sum_{n=\!\!\!\!/\,\, 0} c_n\Bigl(
\varphi_n^2(\sigma,\tau) \theta^2_n - {\rm
i}(\omega_n-k_n)\varphi_n^1(\sigma,\tau)\Pi \theta_n^1\Bigr)\,,
\ee where $\theta^1_0$ and $\theta_0^2$ are fermionic zero modes,
while $\theta_n^1$ and $\theta_n^2$ are fermionic string
oscillators. The coefficients $c_n$ are fixed to be

\be c_n =\frac{1}{\sqrt{1+(\omega_n -k_n)^2}}\,, \qquad n=\pm
1,\pm 2,\ldots \ee With this normalization for the coefficients
$c_n$ the canonical Hamiltonian approach (see Section
\ref{secphaact}) gives the Poisson-Dirac anticommutation relations

\be\label{ttcom1}\{\theta_m^{1\alpha},\theta_n^{1\beta}\}
=\frac{\rm i}{4}(\gamma^+)^{\alpha\beta} \delta_{m+n,0}\,,\qquad
\{\theta_m^{2\alpha},\theta_n^{2\beta}\} =\frac{\rm
i}{4}(\gamma^+)^{\alpha\beta} \delta_{m+n,0}\ee

$$ \qquad m,n=0,\pm1,\pm 2,\ldots\,.$$ We can now quantize our $2d$
string fields $x^I$ and $\theta'$s by promoting the various
Fourier components appearing in
\rf{xIsol},\rf{th1sol},\rf{th2sol} to operators. Supplementing
this by promoting classical (anti)brackets \rf{aacom}\rf{ttcom1}
to equal time (anti)commutators according to the rules $\{.\,,
.\}_{clas}\rightarrow {\rm i} \{.\,, .\}_{quant}$, $[.\,,
.]_{clas}\rightarrow {\rm i} [.\,, .]_{quant}$ we obtain the final
form of the quantum equal-time (anti)commutators. We note that if
we restore dependence on the dimensionful parameter $m$ (see
\rf{lin1}), then in flat space limit ($ m\rightarrow 0$), the
solutions to the equations of motion reduces to the well known
solutions of the flat superstring equations of motion.

\subsection{Superstring action in various coordinates
and Killing \\ parametrization of superspace}\label{kilpar}

In this section we discuss an alternative form of the superstring
Lagrangian that may be useful in various applications. This form
of the Lagrangian is based on the parametrization of plane wave
space given by

\be ds^2\label{lin2} =2dx^+ dx^- +\cos^2 x^+ dx_I^2\,.\ee As is
well known this line interval can be obtained from the one in
\rf{lin1} by replacing the variables as

 \be\label{xtrans} x^- \rightarrow x^- + \frac{1}{4}
x_I^2\sin 2x^+\,, \qquad x^I \rightarrow \cos x^+ x^I \,.\ee With
simultaneous replacing of the fermionic field\footnote{One can
make sure that passing to the new $\theta'$s \rf{thtrans}
corresponds to usage of Killing parametrization of superspace:
$G(x,\theta) =  {\rm g} (\theta) g(x)$.}

\be\label{thtrans} \theta \rightarrow e^{-{\rm i}x^+\Pi} \theta\ee
equations \rf{kin2},\rf{wz2} lead to the following superstring
Lagrangian in fermionic light-cone gauge

\be {\cal L}_{kin}  = -\frac{1}{2}\sqrt{g}g^{ab}\Bigl(2\partial_a
x^+
\partial_b x^-  + \cos^2 x^+\partial_a
x^I \partial_b x^I\Bigr)  -  {\rm i}\sqrt{g}g^{ab}\partial_b
x^+\Bigl( \bar{\theta}\bar{\gamma}^-\partial_a\theta +
\theta\bar{\gamma}^-
\partial_a\bar{\theta}\Bigr)
\ee \be {\cal L}_{WZ} = {\rm i}\epsilon^{ab}\partial_a x^+ \theta
\bar{\gamma}^- e^{-2{\rm i}x^+\Pi}\partial_b \theta +h.c. \ee The
corresponding Lagrangian in the bosonic light-cone gauge can be
obtained using gauge \rf{conf} and fixing the residual conformal
diffeomorphism symmetry as in \rf{xpsol}. The resulting
Lagrangian is

\begin{eqnarray}\label{lag5}
 {\cal L}  & = &  -\frac{1}{2}
\cos^2 x^+ \partial_a x^I
\partial^a x^I \nonumber\\
&
 + & {\rm i} ( \bar{\theta}\bar{\gamma}^-\partial_0\theta +
\theta\bar{\gamma}^-  \partial_0\bar{\theta}+ \theta
\bar{\gamma}^- e^{-2{\rm i}x^+\Pi}\partial_1 \theta + \bar{\theta}
\bar{\gamma}^- e^{2{\rm i}x^+\Pi}\partial_1 \bar{\theta})\,.
 \end{eqnarray}

Compared to \rf{lag4} this  superstring Lagrangian does not
involve mass like terms for superstring $2d$ fields. One can
expect that this parametrization is most convenient for
establishing the connection with the NSR formulation. Solution to
equations of motion obtained from \rf{lag5} can easily be
obtained from \rf{xIsol},\rf{th1sol}\rf{th2sol}  using
transformations \rf{xtrans},\rf{thtrans}.

%%%%%%%%%%%%%%%%%%%%%%%%%%%%%%%%%%%%%%%%%%
\newsection{Light-cone  Hamiltonian approach to
 superstring in plane wave geometry}\label{Phaction}

In this section, following GGRT approach \cite{ggrt}, we develop
phase space formulation of superstring in the plane wave RR
background. This formulation is most convenient for deriving of
Noether charges of basic symmetry superalgebra.

The superstring Lagrangian \rf{ltot2},\rf{kin2}\rf{wz2} can be
represented as

\be\label{lagdec} {\cal L}= {\cal L}_1+{\cal L}_2 \ , \ee where
the two parts are

\begin{eqnarray}
&&
 {\cal L}_1 = -h^{ab} \partial_a x^+ \partial_b x^-
+\frac{1}{2}h^{ab} \partial_a x^+ \partial_b x^+ B +\partial_a
x^+ A^a + C\,,
\\
&& {\cal L}_2 = -\frac{1}{2}h^{ab}\partial_a x^I \partial_b x^I
\end{eqnarray}
and  $h^{ab}$ is defined by

\be \label{hhh} h^{ab}\equiv \sqrt{g}g^{ab}\,, \qquad
h^{00}h^{11}-(h^{01})^2=-1 \ . \ee The functions $A^a$, $B$, and
$C$ take the form

\begin{eqnarray}
\label{afun}&&
 A^a =- {\rm i}
h^{ab}(\bar{\theta}\bar{\gamma}^-\partial_b\theta +
\theta\bar{\gamma}^- \partial_b\bar{\theta} ) + {\rm
i}\epsilon^{a1}(\theta \bar{\gamma}^- \th'
+\bar{\theta}\bar{\gamma}^- \barth') \,,
\\
\label{bfun}&& B = x_I^2 + 4\bar{\theta}\bar{\gamma}^-\Pi
\theta\,,
\\
\label{cfun}&& C =-{\rm i} \x'^+ (\theta\bar{\gamma}^-
\dot{\theta} +\bar{\theta} \bar{\gamma}^-\dot{\bar{\theta}})\,,
\end{eqnarray}
where dot and prime are derivatives over $\tau$ and $\sigma$ (see
Appendix A for  definitions). Decomposition (\ref{lagdec}) is
made so that the functions $A^a$, $B$, $C$, depend  on (i)  the
anticommuting coordinates and their derivatives with respect to
 both  world-sheet coordinates  $\tau$ and $\sigma$ and
(ii)  the  bosonic coordinates and their derivatives with respect
to the world-sheet spatial coordinate $\sigma$ only. The reason
for  this   decomposition is that we use  the
 phase space description with respect to the
bosonic coordinates only,  i.e. we not make the Legendre
transformation with respect to the fermionic coordinates.

%%%%%%%%%%%%%%%%%%%%%%%%%%%%%%%%%%%%%
\subsection{Phase space Lagrangian}\label{secphaact}
%%%%%%%%%%%%%%%%%%%%%%%%

Computing  the canonical momenta for the bosonic coordinates

\be \PP_\mu = \frac{\partial {\cal L}}{\partial \dot{x}^\mu}\,,
\ee we obtain

\begin{eqnarray}
\label{can1}\PP^+ &=& -h^{00}\dot{x}^+ -h^{01}\x'^+\ ,
\\
\PP^I &=& -h^{00}\dot{x}^I -h^{01}\x'^I\ ,
\\
\label{can6} \PP^- &=& -h^{00}\dot{x}^- - h^{01}\x'^- +A^0 -
B\PP^+  \ .
\end{eqnarray}
where we use the conventions $\PP^\pm \equiv \PP_\mp$ and $\PP^I
\equiv \PP_I$. Applying the standard procedure, we then find the
phase space Lagrangian ${\cal L}= {\cal L}_1 + {\cal L}_2$,

\begin{eqnarray}
\label{ll1}{\cal L}_1 &=&
 \PP^+\dot{x}^- + \PP^-\dot{x}^+
+\frac{1}{2h^{00}}\Bigl(2\PP^+\PP^- + 2\x'^+\x'^-  + (\PP^{+2}-
\x'^{+2})B \Bigr)
\nonumber\\
&+& \frac{h^{01}}{h^{00}}(\PP^+\x'^- + \PP^-\x'^+ )
-\frac{1}{h^{00}}(\PP^+ +h^{01}\x'^+)A^0 +\x'^+ A^1 + C\,,
\\
\label{ll2}{\cal L}_2 &=& \PP^I\dot{x}^I
+\frac{1}{2h^{00}}(\PP_I^2 + \x'_I^2) +\frac{h^{01}}{h^{00}}
\PP^I\x'^I\,. \end{eqnarray} We next impose the light-cone gauge

\be\label{lcg} x^+ = \tau \,,\ \ \ \qquad \PP^+=p^+ \ . \ee
Using  these  gauge conditions  in the action and integrating
over $\PP^-$ we arrive at

\be\label{h00} h^{00} = -p^+ \  . \ee Inserting this in
(\ref{ll1}) and (\ref{ll2}) we obtain the general  form of the
 phase space light-cone gauge Lagrangian

\begin{eqnarray} {\cal L}_1 & = & -\frac{1}{2p^+}(p^{+2}B ) -
\frac{h^{01}}{p^+}(p^+\x'^-) +A^0\,,
\\
 {\cal L}_2 & = & \PP^I\dot{x}^I - \frac{1}{2p^+}(\PP_I^2 + \x'_I^2)
-\frac{h^{01}}{p^+}\PP^I\x'^I\,. \end{eqnarray}

 In deriving these  expressions we used that the function
$C$ given in \rf{cfun} is  equal to zero in the light-cone gauge
\rf{lcg}. Now, in order to obtain the light-cone gauge phase
Lagrangian we need only to insert the appropriate functions $A^0$
and $B$ \rf{afun},\rf{bfun}. The result is

\begin{eqnarray} \label{phlag3} {\cal L} &= &
 \PP^I\dot{x}^I + {\rm i}p^+(\bar{\theta}\bar{\gamma}^-\dot{\theta}+
 \theta \bar{\gamma}^- \dot{\bar{\theta}})
 - \frac{1}{2p^+} \Bigl(\PP_I^2 +
\x'_I^2+p^{+2}(x_I^2 + 4\bar{\theta}\bar{\gamma}^-\Pi
\theta)\Bigr)
\nonumber\\
&+&{\rm i}(\theta \bar{\gamma}^-  \th'
+\bar{\theta}\bar{\gamma}^-\barth')
\nonumber\\
&-&\frac{h^{01}}{p^+}\Bigl(p^+\x'^- + \PP^I\x'^I + {\rm
i}p^+(\bar{\theta}\bar{\gamma}^- \th' +\theta \bar{\gamma}^-
\barth' ) \Bigr)\,. \end{eqnarray} This Lagrangian gives the
Hamiltonian

\be\label{ham} P^- =\int d \sigma\   \PP^- \ , \ee where the
Hamiltonian density $\PP^-$ is

\be  \PP^- =
 - \frac{1}{2p^+}\Bigl(\PP_I^2 +
\x'_I^2+p^{+2}(x_I^2 + 4\bar{\theta}\bar{\gamma}^-\Pi
\theta)\Bigr) + {\rm i}(\theta \bar{\gamma}^-  \th'
+\bar{\theta}\bar{\gamma}^-\barth')\,.\ee It should be
supplemented by the  constraint

\be p^+\x'^-  + \PP^I\x'{}^I +{\rm i}p^+(\bar{\theta}
\bar{\gamma}^-\th'+ \theta\bar{\gamma}^- \barth') =0\,. \ee As
usual, this constraint allows one  to express the non-zero modes
of the bosonic
 coordinate $x^-$ in terms of the transverse physical  modes.

The equations of motion corresponding to  the phase space
superstring Lagrangian \rf{phlag3} takes the form

\begin{eqnarray}
&& \dot{x}^I =\frac{1}{p^+}\PP^I \,,
\\
&&
\dot{\PP}^I =\frac{1}{p^+}\x'\!\!\phx{}^I
 -p^+x^I\,,
\\
&&
 \dot{\theta} = - \frac{1}{p^+}\barth' - {\rm i}\Pi\theta\,,
\\
&& \dot{\bar{\theta}} =-\frac{1}{p^+}\th'+{\rm
i}\Pi\bar{\theta}\,.
\end{eqnarray}
These equations can be written in the  Hamiltonian form.
Introducing the notation $\cal X$ for the phase space variables
$(\PP^I,x^I,\theta,\bar{\theta})$ we have the Hamiltonian
equations

\be\label{hamfor} \dot{\cal X} = [{\cal X},\PP^-] \ , \ee where
the phase space variables satisfy the (classical) Poisson-Dirac
brackets

\be\label{pxcom} [\PP^I(\sigma),x^J(\sigma')]
=\delta^{IJ}\delta(\sigma,\sigma')\,, \ee

\be\label{ttcom} \{\bar{\theta}^\alpha(\sigma),
\theta^\beta(\sigma')\} = \frac{\rm
i}{4p^+}(\gamma^+)^{\alpha\beta}\delta(\sigma,\sigma')\,, \ee

\be\label{qcom6}
[x_0^-,\theta^\alpha]=\frac{1}{2p^+}\theta^\alpha, \qquad
[x_0^-,\bar{\theta}^\alpha]=\frac{1}{2p^+}\bar{\theta}^\alpha\,.
\ee Here $x_0^-$ is the  zero mode of $x^-$ and therefore
$[p^+,x_0^-]=1$. All the remaining brackets are equal to zero. To
derive (\ref{ttcom}),(\ref{qcom6}) we must recall that the
Lagrangian \rf{phlag3} has the fermionic
 second class constraints

\be
 p_\alpha+{\rm i}p^+ \gamma_{\alpha\beta}^-\bar{\theta}^\beta=0\,,\qquad
 \bar{p}_\alpha+{\rm i}p^+ \gamma_{\alpha\beta}^-\theta^\beta=0\,,
\qquad \ee
 where $p_\alpha$ and $\bar{p}_\alpha$ are the canonical
momenta of the respective fermionic coordinates $\theta^\alpha$
and $\bar{\theta}^\alpha$. Starting with the Poisson brackets

\be \label{pbr}\{
p_\beta(\sigma),\theta^\alpha(\sigma')\}_{_{P.B.}}
=\frac{1}{2}(\gamma^+\bar{\gamma}^-)^\alpha{}_\beta
\delta(\sigma,\sigma')\,,\qquad
\{\bar{p}_\beta(\sigma),\bar{\theta}^\alpha(\sigma')\}_{_{P.B.}}
=\frac{1}{2}(\gamma^+\bar{\gamma}^-)^\alpha{}_\beta
\delta(\sigma,\sigma')\,,\ee and $[p^+,x_0^-]_{_{P.B.}}=1$ we
then obtain the
 Poisson-Dirac brackets given in (\ref{ttcom}),(\ref{qcom6}).
The projector $(\gamma^+\bar{\gamma}^-)/2$ in (\ref{pbr}) is to
respect light-cone gauge imposed on $\theta^\alpha$ and
$\bar{\theta}^\alpha$ \rf{ferg}.

%%%%%%%%%%%%%%%%%%%%%%%%%%%%%%%%%%%%%%%%%%%%%%%%%%%%%%%
\newsection{Noether charges as generators of  the basic supersymmetry
algebra
 }\label{Ncharges}
%%%%%%%%%%%%%%%%%%%%%%%%%%%%%%%%%%%%%%%%%%%%%%%%%%%%%%%%

The Noether charges play an  important role in analysis of the
symmetries of dynamical systems. The choice of the light-cone
gauge spoils  manifest global symmetries,  and  in order to
demonstrate that these  global invariances are still present, one
needs to  find  the  Noether charges that generate
them.\footnote{In what follows ``currents'' and ``charges'' will
mean both bosonic and fermionic ones, i.e.  will include
supercurrents  and supercharges.} These charges play  a crucial
role  in formulating  superstring   field theory
 in the light-cone gauge  (see \cite{GSB,GS1}).

In  the light-cone formalism, the generators (charges)  of the
basic superalgebra  can be split into two groups:

\begin{equation}\label{kingen}
P^+,\,\, P^I,\,\, J^{+I},\,\, J^{ij},\,\, J^{\ipr\jpr},\,\,
Q^{+},\,\, \bar{Q}^+,
\end{equation}
which we refer to as kinematical generators,  and
\begin{equation}\label{dyngen}
P^-,\,\, Q^-,\,\, \bar{Q}^-,
\end{equation}
which we refer to as dynamical generators. The $Q^+$ and $Q^-$
are expressible in terms of the original supercharges $Q$  as

\be\label{qpqm} Q^+ \equiv \frac{1}{2}\bar{\gamma}^-\gamma^+
Q\,,\qquad
 Q^- \equiv \frac{1}{2}\bar{\gamma}^+\gamma^- Q\,.\ee
For $x^+=0$, the kinematical generators in the superfield
realization are quadratic in the  physical string fields, while
the dynamical generators receive higher-order
interaction-dependent corrections. The first step in  the
construction  of superstring field theory is to find  a  free
(quadratic) superfield representation of the generators of the
basic superalgebra. The charges that we  obtain  below can be used
to obtain  (after quantization) these free superstring field
charges.

%%%%%%%%%%%%%%%%%%%%%%%%
\subsection{Currents for  $\kappa$-symmetry
light-cone  gauge  fixed  superstring action}
%%%%%%%%%%%%%%%%%%%%%%%%%%%

As usual,   symmetry  generating charges   can be obtained from
conserved currents. Because the currents  themselves may be
helpful in some applications, we first derive  them starting with
the $\kappa$-symmetry gauge fixed  Lagrangian in the form given
in \rf{kin2},\rf{wz2}. To obtain the currents, we use the standard
Noether method (see, e.g., \cite{W}) based on the localization of
the parameters of the  associated global transformations. We let
$\epsilon$ be a parameter of some global transformation that
leaves the action  invariant. Replacing it by a function of
world-sheet coordinates $\tau,\sigma$,   the variation of  the
action takes the form

\be\label{dS} \delta S =\int d^2\sigma\  {\cal G}^a
\partial_a \epsilon\, , \ee where ${\cal G}^a$  is the
corresponding current. Using this formula in what follows, we
find those currents that are related to  symmetries that do not
involve compensating $\kappa$-symmetry transformations. The
remaining currents are found  in the next subsection starting from
the action  \rf{phlag3}  where both the $\kappa$-symmetry and the
bosonic  light-cone gauges are fixed.

We start with  the translation invariance. Transformations of the
coordinate under $P^\pm$ translations take the form

 \be
\delta x^\pm = \epsilon^\pm\,, \ee while under $P^I$ translations,
we have the transformations

\be \delta x^- =\sin x^+ \epsilon^I x^I\,,\qquad
 \delta x^I = \cos x^+ \epsilon^I\,.\ee
Applying  (\ref{dS}) to the Lagrangian (\ref{kin2}),(\ref{wz2})
gives the translation currents

\begin{eqnarray}
\PP^{+a} & = & -\sqrt{g}g^{ab}\partial_b x^+\,,
\\
 \PP^{Ia} & = & -\sqrt{g}g^{ab}\cos x^+ \partial_b x^I  +\sin x^+ x^I
\PP^{+a}\,,
\\
\PP^{-a} &=& -\sqrt{g}g^{ab}(\partial_b x^- + {\rm
i}\bar{\theta}\bar{\gamma}^-\partial_b\theta + {\rm
i}\theta\bar{\gamma}^-
\partial_b\bar{\theta})
-x_I^2 \PP^{+a} -  4\PP^{+a}\bar{\theta}\bar{\gamma}^- \Pi \theta
\nonumber\\
&+& ({\rm i}\epsilon^{ab}\theta \bar{\gamma}^-\partial_b \theta
+h.c.)\,.
\end{eqnarray}

Invariance of the action \rf{kin2},\rf{wz2} with respect to
rotations in the $(x^-,x^I)$ plane

\be \delta x^- =\cos x^+ \epsilon^{-I} x^I\,,\qquad
 \delta x^I = -\sin x^+ \epsilon^{-I}\ee
gives the conserved currents

\be {\cal J}^{+I a} = -\sqrt{g}g^{ab} \sin x^+ \partial_b x^I
-\cos x^+ x^I \PP^{+a}\,. \ee

Invariance with respect to the $SO(4)$ and $SO^\prime(4)$
rotations

\be \delta x^i = \epsilon^{ij} x^j\,,\qquad \delta \theta =
\frac{1}{4}\epsilon^{ij}\gamma^{ij}\theta\,,\qquad \delta x^\ipr =
\epsilon^{\ipr\jpr} x^\jpr\,,\qquad \delta \theta =
\frac{1}{4}\epsilon^{\ipr\jpr}\gamma^{\ipr\jpr}\theta\,,\ee
$\epsilon^{ij} = - \epsilon^{ji}$, $\epsilon^{\ipr\jpr} = -
\epsilon^{\jpr\ipr}$, gives the  conserved currents

\begin{eqnarray}
&&\hspace{-1.5cm} {\cal J}^{ij a} = -\sqrt{g}g^{ab}(x^i \partial_b
x^j - x^j\partial_b x^i) -{\rm
i}\bar{\theta}\bar{\gamma}^-\gamma^{ij}\theta \PP^{+a}
-\Bigl(\frac{{\rm i}
\epsilon^{ab}}{2\sqrt{g}}\PP_b^+\theta\bar{\gamma}^-\gamma^{ij}\theta
+h.c.\Bigr)\,,
\\
&&\hspace{-1.5cm}
 {\cal J}^{\ipr\jpr a} =-\sqrt{g}g^{ab}(x^\ipr \partial_b x^\jpr
- x^\jpr\partial_b x^\ipr) -{\rm
i}\bar{\theta}\bar{\gamma}^-\gamma^{\ipr\jpr}\theta \PP^{+a}
-\Bigl(\frac{{\rm i}
\epsilon^{ab}}{2\sqrt{g}}\PP_b^+\theta\bar{\gamma}^-\gamma^{\ipr\jpr}\theta
+h.c.\Bigr),\end{eqnarray}
 where $\PP_a^+\equiv g_{ab}\PP^{+b}$.

Invariance with respect to the super transformations

\be \delta \theta = e^{-{\rm i}x^+\Pi} \epsilon\,,\qquad \delta
\bar{\theta} = e^{{\rm i}x^+\Pi} \bar{\epsilon}\,,\qquad
 \delta x^- = -{\rm i}\epsilon \bar{\gamma}^- e^{-{\rm
i}x^+\Pi}\bar{\theta} -{\rm i}\bar{\epsilon} \bar{\gamma}^-
e^{{\rm i}x^+\Pi}\theta\,,\ee leads to the conserved supercurrents

 \be {\cal Q}^{+a} = 2\bar{\gamma}^-e^{{\rm i}x^+
\Pi} (\PP^{+a}\theta
+\frac{\epsilon^{ab}}{\sqrt{g}}\PP^+_b\bar{\theta})\,,\ee

 \be \bar{{\cal Q}}^{+a} = 2\bar{\gamma}^-e^{-{\rm i}x^+
\Pi} (\PP^{+a}\bar{\theta}
+\frac{\epsilon^{ab}}{\sqrt{g}}\PP^+_b\theta)\,.\ee

\subsection{Charges for  bosonic and $\kappa$-symmetry
  light-cone gauge fixed
superstring action}
%%%%%%%%%%%%%%%%%%%%%%%%%%%%%%%%%%

In  the previous section, we have found (super)currents starting
with the  $\kappa$-symmetry light-cone gauge fixed action given
in \rf{kin2},\rf{wz2}. These currents can be used  to find
currents for  the  action where both the fermionic
$\kappa$-symmetry and the bosonic  reparametrization symmetry are
fixed by light-cone gauges \rf{phlag3}. To find the components of
the currents in the world-sheet time direction ${\cal G}^0$ we
should use relations  for the canonical momenta
\rf{can1}--\rf{can6} and then to insert the  light-cone gauge
conditions (\ref{lcg}) and (\ref{h00}) in the expressions for the
currents given in the previous subsection. The  charges are then
given by

\be G =\int d\sigma\  {\cal G}^0 \ . \ee We start with the
kinematical generators (charges) (\ref{kingen}). The results for
the currents imply the following representations for some of them

\be\label{kin1} P^+  = p^+ \,,\qquad
 P^I = \int \cos x^+ \PP^I  +\sin x^+ x^I p^+\,,
\ee

\be J^{+I} = \int  \sin x^+ \PP^I -\cos x^+ x^I p^+ \,,\ee

 \be \label{kin3}  Q^+ = \int 2p^+\bar{\gamma}^-e^{{\rm i}x^+
\Pi}\theta \,,\qquad
 \bar{Q}^+ = \int 2p^+\bar{\gamma}^-e^{-{\rm i}x^+
\Pi}\bar{\theta}\,. \ee We note that these charges depend only on
the zero modes of string coordinates. In \rf{kin1}--\rf{kin3} the
integrands are ${\cal G}^0$ parts of the corresponding currents
in world-sheet time direction: ${\cal P}^{I0}$, ${\cal J}^{+ I
0}$, ${\cal Q}^{+0}$, $\bar{{\cal Q}}^{+ 0}$, and ${\cal
P}^{+0}=p^+$.

The remaining kinematical charges depend on non-zero string modes
and are given by

\be  J^{ij} = \int x^i \PP^j - x^j\PP^i  -{\rm
i}p^+\bar{\theta}\bar{\gamma}^-\gamma^{ij}\theta \,, \ee

\be J^{\ipr\jpr } = \int x^\ipr \PP^\jpr - x^\jpr\PP^\ipr  -{\rm
i}p^+\bar{\theta}\bar{\gamma}^-\gamma^{\ipr\jpr}\theta\,. \ee

The dynamical charge $P^-$ is given by \rf{ham}, while the
supercharges $Q^-$ and $\bar{Q}^-$ are given by

\be \label{dyncha1} Q^{-} = \int 2\PP^I\bar{\gamma}^I\theta -
2\x'{}^I\bar{\gamma}^I \bar{\theta} + 2{\rm i}p^+
x^I\bar{\gamma}^I \Pi\theta \,,\ee

\be  \bar{Q}^{-} = \int 2\PP^I\bar{\gamma}^I\bar{\theta} -
2\x'{}^I\bar{\gamma}^I \theta - 2{\rm i}p^+ x^I\bar{\gamma}^I
\Pi\bar{\theta}\,. \ee The derivation of these supercharges can be
found in Appendix B.

%%%%%%%%%%%%%%%%%%%%%%%%%%
\newsection{Conclusions}\label{concl}
%%%%%%%%%%%%%%%%%%%%%%%%%

We have developed the $\kappa$-symmetric and  light-cone gauge
formulations of type $IIB$ superstring in the plane wave
Ramond-Ramond background. We restricted our consideration to the
study of classical superstring dynamics. Because in light-cone
gauge the superstring action is quadratic in $2d$ fields, this
superstring model can be explicitly quantized in a rather
straightforward way. We have presented various forms of the
light-cone gauge superstring Lagrangian. In coordinates \rf{lin1}
the light-cone gauge Lagrangian given by \rf{lag2dform} (or
\rf{lag4}) describes 8 massive bosonic $2d$ scalars and 8 massive
Majorana $2d$ fermions propagating in flat $2d$ world-sheet. In
coordinates \rf{lin2} the light-cone gauge Lagrangian given by
\rf{lag5} describes 8 massless bosonic and 8+8 fermionic $2d$
superstring fields. The fact that the light-cone gauge action is
quadratic in
 physical fields implies that the corresponding plane wave background
with RR 5-flux should be an {\it exact} string solution.

As noted above, the relatively simple model of superstring in the
plane wave RR background  may serve as a training ground for the
study of the more interesting case of superstring in $AdS_5\times
S^5$. Because the plane wave superstring action is invariant
under the global transformations of the basic symmetry
superalgebra of the plane wave RR background, the string spectrum
should be classified by unitary representations of this
superalgebra. These unitary representations appear as Fourier
modes of solutions to free equations of motion for quantized
fields propagating in the plane wave RR background. The fields in
the plane wave background have the following two features in
common with the fields in the $AdS$ space time: (i) The spectrum
of the energy operator is discrete; (ii) The spectrum of the
energy operator even for massless representations is bounded from
below by a {\it nonzero} value \cite{metpl}. From this line of
reasoning we believe that the study of the plane wave RR
superstring will be useful for better understanding  strings in
AdS/RR-charge backgrounds. We note that in this paper we have
developed GS formulation of plane wave/RR superstring. NSR
formulation of this superstring  is still to be understood.

%%%%%%%%%%%%%%%%%%%%%%%%%%%%%%%%%%%%%%%%%%%%%%%%%
\section*{Acknowledgments}
%%%%%%%%%%%%%%%%%%%%%%%%%%%%%%%%%%%%%%%%%%%%%%%%%%

Author would like to thank A. Semikhatov for reading the
manuscript and useful comments. This work was supported by the
INTAS project 00-00254, by the RFBR Grant No.99-02-17916, and
RFBR Grant for Leading Scientific Schools, Grant No. 01-02-30024.

%%%%%%%%%%%%%%%%%%%%%%%%%%%%%%%%%%%%%%%%%%%%%%%%
\setcounter{section}{0} \setcounter{subsection}{0}
%%%%%%%%%%%%%%%%%%%%%%%%%%%%%%%%%%%%%%%%%%%%%%%%%

\appendix{Notation}
\label{not}

In the main part of the paper, we use the following  conventions
for the indices:
\begin{eqnarray*}
\mu,\nu,\rho = 0,1,\ldots, 9 && \qquad  so(9,1) \  \hbox{ vector
indices (tangent space indices) }
\\
I,J,K,L = 1,\ldots, 8 && \qquad  so(8) \  \hbox{ vector indices
(tangent space indices) }
\\
i,j,k,l = 1,\ldots, 4 && \qquad  so(4) \  \hbox{ vector indices
(tangent space indices) }
\\
\ipr,\jpr,\kpr,\lpr = 5,\ldots, 8 && \qquad  so^\prime(4) \
\hbox{ vector indices (tangent space indices) }
\\
\alpha,\beta,\gamma = 1,\ldots, 16 && \qquad  so(9,1) \  \hbox{
spinor indices in chiral representation}
\\
a,b = 0,1 && \qquad 2d \hbox{ world-sheet coordinate indices}
\end{eqnarray*}
We suppress the flat space metric tensor
$\eta_{\mu\nu}=(-,+,\ldots, +)$ in scalar products, i.e.

\be X^\mu Y^\mu\equiv \eta_{\mu\nu}X^\mu Y^\nu \ee We decompose
$x^\mu$ into the light-cone and transverse coordinates: $x^\mu=
(x^+,x^-, x^I)$, $x^I=(x^i,x^\ipr)$, where

\be x^\pm\equiv \frac{1}{\sqrt{2}}(x^9\pm x^0)\,.\ee In this
notation, scalar products of tangent space vectors are decomposed
as

\be X^\mu Y^\mu = X^+Y^- + X^-Y^+ +X^IY^I\,, \qquad
X^IY^I=X^iY^i+X^\ipr Y^\ipr\,. \ee Unless stated otherwise, we
always assume the summation over repeated indices (irrespective
of their position).

The  derivatives with respect to the  world-sheet coordinates
$(\tau,\sigma)$ are

\be \dot{x}^I \equiv \partial_\tau x^I\,, \qquad \x'^I \equiv
\partial_\sigma x^I\,. \ee

The world-sheet Levi-Civita symbol $\epsilon^{ab}$   is defined
with $\epsilon^{01}=1$.

We use the chiral representation for the $32\times 32$ Dirac
matrices $\Gamma^\mu$ in terms of the $16\times 16 $ gamma
matrices $\gamma^\mu$

\be \Gamma^\mu =\left(\begin{array}{cc} 0  & \gamma^\mu \\
\bar{\gamma}^\mu & 0
\end{array}\right) \,,\ee

\be \gamma^\mu\bar{\gamma}^\nu + \gamma^\nu\bar{\gamma}^\mu
=2\eta^{\mu\nu}\,,\qquad \gamma^\mu =
(\gamma^\mu)^{\alpha\beta}\,, \qquad  \bar{\gamma}^\mu
=\gamma^\mu_{\alpha\beta}\,, \ee

\be \gamma^\mu=(1,\gamma^I,\gamma^9)\,,\qquad
\bar{\gamma}^\mu=(-1,\gamma^I,\gamma^9)\,,\qquad
\alpha,\beta=1,\ldots 16\,.\ee We adopt the Majorana
representation for $\Gamma$-matrices, $ C= \Gamma^0$, which
implies that all $\gamma^\mu$ matrices are real and symmetric,
$\gamma^\mu_{\alpha\beta} = \gamma^\mu_{\beta\alpha}$,
$(\gamma^\mu_{\alpha\beta })^* = \gamma^\mu_{\alpha\beta}$.

We use the convention that $\gamma^{\mu_1\ldots \mu_k}$ are the
antisymmetrized product of $k$ gamma matrices normalized so that

\be (\gamma^{\mu\nu})^\alpha{}_\beta \equiv
\frac{1}{2}(\gamma^\mu\bar{\gamma}^\nu)^\alpha{}_\beta -(\mu
\leftrightarrow \nu)\,,\qquad
(\bar{\gamma}^{\mu\nu})_\alpha{}^\beta \equiv
\frac{1}{2}(\bar{\gamma}^\mu \gamma^\nu)_\alpha{}^\beta -(\mu
\leftrightarrow \nu)\,.\ee

\be (\gamma^{\mu\nu\rho})^{\alpha\beta} \equiv
\frac{1}{6}(\gamma^\mu\bar{\gamma}^\nu\gamma^\rho)^{\alpha\beta}
\pm 5 \hbox{ terms} \,,\qquad
(\bar{\gamma}^{\mu\nu\rho})_{\alpha\beta} \equiv
\frac{1}{6}(\bar{\gamma}^\mu
\gamma^\nu\bar{\gamma}^\rho)_{\alpha\beta} \pm 5\hbox{
terms}\,.\ee We assume the normalization $ \gamma^0\bar{\gamma}^1
\ldots \gamma^8\bar{\gamma}^9=1$, i.e.

\be\label{g11} \Gamma_{11} \equiv  \Gamma^0\ldots \Gamma^9
=\left(\begin{array}{cc}
1 & 0\\
0 & -1
\end{array}\right)\,.
\ee We use the following notation $\Pi = \Pi^\alpha{}_\beta$,
$\Pi^\prime = (\Pi^\prime)^\alpha{}_\beta$ where

\be \Pi^\alpha{}_\beta \equiv
(\gamma^1\bar{\gamma}^2\gamma^3\bar{\gamma}^4)^\alpha{}_\beta\,,\qquad
 (\Pi^\prime)^\alpha{}_\beta \equiv
(\gamma^5\bar{\gamma}^6\gamma^7\bar{\gamma}^8)^\alpha{}_\beta\,.\ee
Because of the relation $\gamma^0\bar{\gamma}^9 =\gamma^{+-}$ the
normalization condition \rf{g11} takes the form

 \be \gamma^{+-}\Pi\Pi^\prime = 1\,.\ee
We have the relations

\be (\gamma^{+-})^2 = \Pi^2 =(\Pi^\prime)^2 =1\,,\ee

\be \gamma^{+-}\gamma^{\pm} =\pm \gamma^\pm\,,\qquad
\bar{\gamma}^\pm \gamma^{+-} = \mp \bar{\gamma}^\pm\,,\qquad
\gamma^+\bar{\gamma}^+ = \gamma^-\bar{\gamma}^- =0\,,\ee

\be \bar{\gamma}^+(\Pi+\Pi^\prime)=(\Pi+\Pi^\prime) \gamma^-
=0\,, \qquad \bar{\gamma}^-(\Pi - \Pi^\prime)= (\Pi - \Pi^\prime)
\gamma^+ =0\,. \ee The 32-component positive chirality $\theta$
and negative chirality $Q$ spinors are decomposed in terms of the
16-component spinors as

\be \theta = \left( \begin{array}{c} \theta^\alpha \\
0\end{array}\right)\,, \qquad\quad
Q = \left( \begin{array}{c} 0 \\
Q_\alpha\end{array}\right)\,.\ee Instead of  {\it one complex}
Weyl spinor $\theta$, we sometimes prefer to use {\it two real}
Majorana-Weyl spinors $\theta^1$ and $\theta^2$ defined by

\be\label{comrea} \theta = \frac{1}{\sqrt{2}}(\theta^1 +{\rm
i}\theta^2)\,,\qquad \bar{\theta} = \frac{1}{\sqrt{2}}(\theta^1 -
{\rm i}\theta^2)\,.\ee We use the shorthand notation like
$\bar{\theta}\bar{\gamma}^\mu\theta$ and $\bar{\gamma}^\mu\theta$
which should read as
$\bar{\theta}{}^\alpha\gamma_{\alpha\beta}^\mu\theta^\beta$ and
$\gamma_{\alpha\beta}^\mu \theta^\beta$ respectively.

\appendix{Derivation of supercharges}

%%%%%%%%%%%%%%%%%%%%%%%%%%%%%%%%%%%%%%%%%%%%%%%%%%%%

Here we demonstrate how the knowledge  of the kinematical charges
and commutation relations of the superalgebra allows one to
obtain the dynamical supercharges $Q^-$ systematically. Before
proceeding we write down the (anti)commutation relation of the
basic algebra \rf{jq1}-\rf{qq} in terms of the $Q^+$ and $Q^-$
supercharges defined in \rf{qpqm}. Using \rf{qpqm} in
\rf{jq1}-\rf{qq} we have

\be \label{jijqapp}[J^{ij},Q_\alpha^\pm] = \frac{1}{2}Q_\beta^\pm
(\gamma^{ij})^\beta{}_\alpha \,, \qquad
[J^{\ipr\jpr},Q_\alpha^\pm] =
\frac{1}{2}Q_\beta^\pm(\gamma^{\ipr\jpr})^\beta{}_\alpha  \ee

 \be\label{jq3} [J^{+I},Q_\alpha^-] =
 \frac{1}{2}Q_\beta^+(\gamma^{+I})^\beta{}_\alpha
\ee

\be\label{pq3} [P^I,Q_\alpha^-] =  \frac{\rm i}{2} Q_\beta^+( \Pi
\gamma^{+I})^\beta{}_\alpha\,, \qquad [P^-,Q_\alpha^+] =  {\rm i}
Q_\beta^+\Pi^\beta{}_\alpha\,,
 \ee

\begin{eqnarray} \label{qpqpapp}
&&
\{Q_\alpha^+,\bar{Q}_\beta^+\} = -2{\rm
i}\gamma^-_{\alpha\beta}P^+\,,\\
&& \{Q_\alpha^+,\bar{Q}_\beta^-\} =-{\rm
i}(\bar{\gamma}^-\gamma^+\bar{\gamma}^I)_{\alpha\beta}P^I
-(\bar{\gamma}^-\gamma^+\bar{\gamma}^i\Pi)_{\alpha\beta}J^{+i}
-(\bar{\gamma}^-\gamma^+\bar{\gamma}^\ipr\Pi^\prime)_{\alpha\beta}J^{+\ipr}\,,\\
\label{qmqmapp}&&
 \{Q_\alpha^-,\bar{Q}_\beta^-\}= -2{\rm i}\gamma^+_{\alpha\beta}
P^- +(\bar{\gamma}^+\gamma^{ij}\Pi)_{\alpha\beta}J^{ij}
+(\bar{\gamma}^+\gamma^{\ipr\jpr}\Pi^\prime)_{\alpha\beta}J^{\ipr\jpr}\,.
\end{eqnarray}

We now let ${\cal Q}^{-a}$ be the conserved supercurrent whose
component in the world-sheet time direction gives the supercharge
$Q^- = \int {\cal Q}^{-0}$. We start with the anzats

\be {\cal Q}^{-0} = \PP^IA_1^I\theta + x^IA_2^I\theta +\x' B^I
\bar{\theta}\,,\ee where the coefficients $A_1^I$, $A_2^I$, and
$B^I$ are assumed to be independent of the dynamical variables
$x^I$, $\PP^I$, $\theta$, $\bar{\theta}$. Using the commutator
\rf{jq3}, first commutator in \rf{pq3}, and Poissoin-Dirac
brackets \rf{pxcom},\rf{ttcom} we obtain the following equations

\be p^+A_1^I\cos x^+ + A_2^I\sin x^+ = 2p^+ \bar{\gamma}^I e^{{\rm
i}x^+\Pi}\,,\ee

\be -p^+A_1^I\sin x^+ + A_2^I\cos x^+ = 2{\rm i}p^+
\bar{\gamma}^I\Pi e^{{\rm i}x^+\Pi}\,.\ee The solution to these
equations is found to be

\be A_1^I =2\bar{\gamma}^I\,,\qquad A_2^I = 2{\rm
i}p^+\bar{\gamma}^I\Pi\,.\ee Thus we have

\be {\cal Q}^{-0} = 2\PP^I\bar{\gamma}^I\theta + 2{\rm i}p^+ x^I
\bar{\gamma}^I \Pi\theta  + \x' B^I \bar{\theta}\,.\ee The
conservation law for the supercurrent ${\cal Q}^{-a}$

\be
\partial_0 {\cal Q}^{-0} +\partial_1 {\cal Q}^{-1}=0\ee
gives $B^I = -2\bar{\gamma}^I$ and fixes the ${\cal Q}^{-0}$ and
${\cal Q}^{-1}$ to be

\be {\cal Q}^{-0} = 2\PP^I\bar{\gamma}^I\theta -
2\x'{}^I\bar{\gamma}^I \bar{\theta} + 2{\rm i}p^+
x^I\bar{\gamma}^I \Pi\theta \,,\ee

\be {\cal Q}^{-1} = \frac{2}{p^+}\PP^I\bar{\gamma}^I\bar{\theta}
-\frac{2}{p^+}\x'{}^I\bar{\gamma}^I \theta + 2{\rm
i}x^I\bar{\gamma}^I \Pi\bar{\theta}\,. \ee The expression for
${\cal Q}^{-0}$ leads to the supercharges $Q^-$ given by
\rf{dyncha1}. Finally, we check that the supercharges satisfy all
the remaining (anti)commutation relation of the basic
superalgebra. The corresponding (anti)commutation relations can
be obtained from the \rf{jijqapp}-\rf{qmqmapp} by changing signs
in anticommutators $\{.\,.\}\rightarrow - \{.\,.\}$ there.


\newpage
\begin{thebibliography}{30}
\parskip-2pt

%\cite{Schwarz:qr}
\bibitem{schwarz}
J.~H.~Schwarz,
%``Covariant Field Equations Of Chiral N=2 D = 10 Supergravity,''
Nucl.\ Phys.\ B {\bf 226}, 269 (1983).
%%CITATION = NUPHA,B226,269;%%

%\cite{Green:1982tk}
\bibitem{gs0}
M.~B.~Green and J.~H.~Schwarz,
%``Extended Supergravity In Ten-Dimensions,''
Phys.\ Lett.\ B {\bf 122}, 143 (1983).
%%CITATION = PHLTA,B122,143;%%

%\cite{Schwarz:wa}
\bibitem{sw}
J.~H.~Schwarz and P.~C.~West,
%``Symmetries And Transformations Of Chiral
%N=2 D = 10 Supergravity,''
Phys.\ Lett.\ B {\bf 126}, 301 (1983).
%%CITATION = PHLTA,B126,301;%%


%\cite{Blau:2001ne}
\bibitem{blau}
M.~Blau, J.~Figueroa-O'Farrill, C.~Hull and G.~Papadopoulos, ``A
new maximally supersymmetric background of IIB superstring
theory,'' arXiv:hep-th/0110242.
%%CITATION = HEP-TH 0110242;%%



%\cite{Kowalski-Glikman:ux}
\bibitem{kg1}
J.~Kowalski-Glikman,
%``A Nontrivial Vacuum State In D = 10, N=1 Supergravity,''
Phys.\ Lett.\ B {\bf 139}, 159 (1984).
%%CITATION = PHLTA,B139,159;%%

%\cite{Chrusciel:1984gr}
\bibitem{kg2}
P.~T.~Chrusciel and J.~Kowalski-Glikman,
%``The Isometry Group And Killing
%Spinors For The PP Wave Space-Time
%In D = 11 Supergravity,''
Phys.\ Lett.\ B {\bf 149}, 107 (1984).
%%CITATION = PHLTA,B149,107;%%

%\cite{Kallosh:1998qs}
\bibitem{kraj}
R.~Kallosh and A.~Rajaraman,
%``Vacua of M-theory and string theory,''
Phys.\ Rev.\ D {\bf 58}, 125003 (1998) [arXiv:hep-th/9805041].
%%CITATION = HEP-TH 9805041;%%


%\cite{Metsaev:1998it}
\bibitem{mt1}
R.~R.~Metsaev and A.~A.~Tseytlin,
%``Type IIB superstring action in AdS(5) x S(5) background,''
Nucl.\ Phys.\ B {\bf 533}, 109 (1998) [arXiv:hep-th/9805028].
%%CITATION = HEP-TH 9805028;%%


%\cite{Kallosh:1998zx}
\bibitem{krr}
R.~Kallosh, J.~Rahmfeld and A.~Rajaraman,
%``Near horizon superspace,''
JHEP {\bf 9809}, 002 (1998) [arXiv:hep-th/9805217].
%%CITATION = HEP-TH 9805217;%%


%\cite{Metsaev:1998hf}
\bibitem{mt2}
R.~R.~Metsaev and A.~A.~Tseytlin,
%``Supersymmetric D3 brane action in AdS(5) x S**5,''
Phys.\ Lett.\ B {\bf 436}, 281 (1998) [arXiv:hep-th/9806095].
%%CITATION = HEP-TH 9806095;%%


%\cite{Metsaev:2000yf}
\bibitem{mt3}
R.~R.~Metsaev and A.~A.~Tseytlin,
%``Superstring action in AdS(5) x S(5):
%kappa-symmetry light-cone gauge,''
Phys.\ Rev.\ D {\bf 63}, 046002 (2001) [arXiv:hep-th/0007036].
%%CITATION = HEP-TH 0007036;%%


%\cite{Metsaev:2000yu}
\bibitem{mtt}
R.~R.~Metsaev, C.~B.~Thorn and A.~A.~Tseytlin,
%``Light-cone superstring in AdS space-time,''
Nucl.\ Phys.\ B {\bf 596}, 151 (2001) [arXiv:hep-th/0009171].
%%CITATION = HEP-TH 0009171;%%

%\cite{Grisaru:fv}
\bibitem{ghm}
M.~T.~Grisaru, P.~Howe, L.~Mezincescu, B.~Nilsson and
P.~K.~Townsend,
%``N=2 Superstrings In A Supergravity Background,''
Phys.\ Lett.\ B {\bf 162}, 116 (1985).
%%CITATION = PHLTA,B162,116;%%


%\cite{Howe:sr}
\bibitem{hw}
P.~S.~Howe and P.~C.~West,
%``The Complete N=2, D = 10 Supergravity,''
Nucl.\ Phys.\ B {\bf 238}, 181 (1984).
%%CITATION = NUPHA,B238,181;%%

%\cite{Zhou:1999sm}
\bibitem{zho}
J.~G.~Zhou,
%``Super 0-brane and GS superstring actions on AdS(2) x S(2),''
Nucl.\ Phys.\ B {\bf 559}, 92 (1999) [arXiv:hep-th/9906013].
%%CITATION = HEP-TH 9906013;%%


\bibitem{pes2} I.~Pesando,
%``The GS type IIB superstring action
%on $AdS_3\times S^3\times T^4$,''
JHEP {\bf 02}, 007 (1999), hep-th/9809145.
%%CITATION = JHEPA,9902,007;%%

\bibitem{ramraj} J.~Rahmfeld and A.~Rajaraman,
%``The GS string action on $AdS_3\times S^3$ with Ramond-Ramond charge,''
Phys.\ Rev.\ {\bf D60}, 064014 (1999), hep-th/9809164.
%%CITATION = PHRVA,D60,064014;%%

\bibitem{parrey} J.~Park and S.~Rey,
%``Green-Schwarz superstring on $AdS_3 \times S^3$,''
JHEP {\bf 01}, 001 (1999), hep-th/9812062.
%%CITATION = JHEPA,9901,001;%%

%\cite{Dawson:2000tw}
\bibitem{daw}
P.~Dawson,
%``D1 and D5-brane actions in AdS(m) x S(n),''
Int.\ J.\ Mod.\ Phys.\ A {\bf 16}, 267 (2001)
[arXiv:hep-th/0002030].
%%CITATION = HEP-TH 0002030;%%

%\cite{deWit:1998yu}
\bibitem{wpps}
B.~de Wit, K.~Peeters, J.~Plefka and A.~Sevrin,
%``The M-theory two-brane in AdS(4) x S(7) and AdS(7) x S(4),''
Phys.\ Lett.\ B {\bf 443}, 153 (1998) [arXiv:hep-th/9808052].
%%CITATION = HEP-TH 9808052;%%


%\cite{Pasti:1999sp}
\bibitem{Pasti:1999sp}
P.~Pasti, D.~Sorokin and M.~Tonin, ``Branes in super-AdS
backgrounds and superconformal theories,'' Talk given at BLTP
International Workshop on Supersymmetry and Quantum Symmetries,
Dubna, Russia, 26-31 Jul 1999. arXiv:hep-th/9912076.
%%CITATION = HEP-TH 9912076;%%

%\cite{Delduc:2001tb}
\bibitem{dik}
F.~Delduc, E.~Ivanov and S.~Krivonos,
%``Partial supersymmetry breaking and AdS(4) supermembrane,''
arXiv:hep-th/0111106.
%%CITATION = HEP-TH 0111106;%%

\bibitem{ggrt} P.~Goddard, J.~Goldstone, C.~Rebbi and C.B.~Thorn,
%``Quantum Dynamics Of A Massless Relativistic String,''
Nucl.\ Phys.\  {\bf B56}, 109 (1973).
%%CITATION = NUPHA,B56,109;%%

%\cite{Forgacs:1995tx}
\bibitem{forhor}
P.~Forgacs, P.~A.~Horvathy, Z.~Horvath and L.~Palla,
%``The Nappi-Witten string in the light-cone gauge,''
Heavy Ion Phys.\  {\bf 1}, 65 (1995) [arXiv:hep-th/9503222].
%%CITATION = HEP-TH 9503222;%%


%\cite{vanNieuwenhuizen:ke}
\bibitem{vann}
P.~van Nieuwenhuizen, ``General Theory Of Coset Manifolds And
Antisymmetric Tensors Applied To Kaluza-Klein Supergravity,''
ITP-SB-84-57 {\it  In *Trieste 1984, Proceedings, Supersymmetry
and Supergravity '84*, 239-323}.

%\cite{Green:1983wt}
\bibitem{gs}
M.~B.~Green and J.~H.~Schwarz,
%``Covariant Description Of Superstrings,''
Phys.\ Lett.\ B {\bf 136}, 367 (1984).
%%CITATION = PHLTA,B136,367;%%
%\cite{Green:1983sg}
%``Properties Of The Covariant
%Formulation Of Superstring Theories,''
Nucl.\ Phys.\ B {\bf 243}, 285 (1984).
%%CITATION = NUPHA,B243,285;%%


%\cite{Henneaux:mh}
\bibitem{hm}
M.~Henneaux and L.~Mezincescu,
%``A Sigma Model Interpretation Of Green-Schwarz
%Covariant Superstring Action,''
Phys.\ Lett.\ B {\bf 152}, 340 (1985).
%%CITATION = PHLTA,B152,340;%%


%\cite{Siegel:1983hh}
\bibitem{ws}
W.~Siegel,
%``Hidden Local Supersymmetry In The
%Supersymmetric Particle Action,''
Phys.\ Lett.\ B {\bf 128}, 397 (1983).
%%CITATION = PHLTA,B128,397;%%

\bibitem{pes} I.~Pesando,
%``A kappa gauge fixed type IIB superstring
%action on $AdS_5 \times S^5$,''
JHEP {\bf 9811}, 002 (1998),
[hep-th/9808020];
%%CITATION = HEP-TH 9808020;%%
%``All roads lead to Rome: Supersolvable and supercosets,''
Mod.\ Phys.\ Lett.\  {\bf A14}, 343 (1999) [hep-th/9808146].

\bibitem{kr} R.~Kallosh and J.~Rahmfeld,
%``The GS string action on $AdS_5 \times S^5$,''
Phys.\ Lett.\ {\bf B443}, 143 (1998),
hep-th/9808038.
%%CITATION = HEP-TH 9808038;%%

\bibitem{bdh}
 L.~Brink, P.~Di Vecchia and P.~Howe,
%``A Locally Supersymmetric And Re\-pa\-ra\-metrization Invariant
%Action For The Spinning String,''
Phys.\ Lett.\  {\bf B65}, 471
(1976).
%%CITATION = PHLTA,B65,471;%%

\bibitem{poly}
 A.M.~Polyakov,
%``Quantum geometry of bosonic strings,''
Phys.\ Lett.\  {\bf B103}, 207 (1981).
%%CITATION = PHLTA,B103,207;%%

\bibitem{GSB}
M.B.~Green, J.~H.~Schwarz and L.~Brink,
%``Superfield Theory Of Type II Superstrings,''
Nucl.\ Phys.\  {\bf B219}, 437 (1983).
%%CITATION = NUPHA,B219,437;%%


\bibitem{GS1}
M.B.~Green and J.~H.~Schwarz,
%``Superstring Field Theory,''
Nucl.\ Phys.\  {\bf B243}, 475 (1984);
%%CITATION = NUPHA,B243,475;%%
%\bibitem{GS2}
%M.B.~Green and J.~H.~Schwarz,
%``Superstring Interactions,''
Nucl.\ Phys.\  {\bf B218}, 43
(1983).
%%CITATION = NUPHA,B218,43;%%


%\cite{Weinberg:1995mt}
\bibitem{W}
S.~Weinberg, ``The Quantum theory of fields. Vol. 1:
Foundations,'' {\it  Cambridge, UK: Univ. Pr.} (1995) 609 p.


%\cite{Metsaev:1997ut}
\bibitem{metpl}
R.~R.~Metsaev,
%``Massless fields in plane wave geometry,''
J.\ Math.\ Phys.\  {\bf 38}, 648 (1997) [arXiv:hep-th/9701141].
%%CITATION = HEP-TH 9701141;%%





\end{thebibliography}
\end{document}